\documentclass[%
preprint,
 amsmath,amssymb,
 aps,showkeys,
]{revtex4-1}
\usepackage[breaklinks,colorlinks=true]{hyperref}
\usepackage[pdftex]{graphicx}
\pdfoutput=1
\usepackage{color}
\usepackage{enumerate}
\usepackage{multirow}

\begin{document}

\preprint{J. R. Soc. Interface 16: 20180814 (2019).\phantom{z}\\
 {\url{ http://dx.doi.org/10.1098/rsif.2018.0814}}}

\title{\bf Group formation on a small-world: experiment and modelling}

\author{Kunal Bhattacharya}
 \email{kunal.bhattacharya@aalto.fi}
\author{Tuomas Takko}
 \email{tuomas.takko@aalto.fi}
\author{Daniel Monsivais}
 \email{daniel.monsivais-velazquez@aalto.fi}
\author{Kimmo Kaski}
 \email{kimmo.kaski@aalto.fi}
\affiliation {Department of Computer Science, Aalto University School of Science, P.O. Box 15400, FI-00076 AALTO, Finland}

\begin{abstract}
As a step towards studying human-agent collectives we conduct an online game with human participants cooperating on a network. The game is presented in the context of achieving group formation through local coordination. The players set initially to a small world network with limited information on the location of other players, coordinate their movements to arrange themselves into groups. To understand the decision making process we construct a data-driven model of agents based on probability matching. The model allows us to gather insight into the nature and degree of rationality employed by the human players. By varying the parameters in agent based simulations we are able to benchmark the human behaviour. We observe that while the players utilize the neighbourhood information in limited capacity, the perception of risk is optimal. We also find that for certain parameter ranges the agents are able to act more efficiently when compared to the human players. This approach would allow us to simulate the collective dynamics in games with agents having varying strategies playing alongside human proxies.
\end{abstract}

\maketitle


\section{\label{sec:level1}Introduction}
Understanding cooperation and conflict in digitized societies is becoming increasingly important with the introduction of artificial intelligence or agents as actors in human social networks. Human-agent cooperation has been lately realized in diverse scenarios ranging from health care settings, to retail stores, to self-driving cars \cite{jennings2014human,van2017domo}. The task of forging cooperation or coordination between humans and agents in complex environments requires the understanding at different levels, such as the psychological underpinnings of human preferences, as well as anticipating the dynamics from human-agent collective action \cite{bonabeau2002agent,kahneman2003maps,groom2007can}. While the answers are far from simple, the use of online games to study hybrid human-agent systems seems to provide valuable insights \cite{shirado2017locally,crandall2018cooperating}. For example, Shirado and Christakis \cite{shirado2017locally} studied how the collective performance of humans trying to solve a coordination game on a network changes in the presence of agents (or bots), and showed the impact of the degree of randomness in agents' behaviour on the outcomes. 

In studying the dynamics of human-agent collectives in games, one needs to investigate the dynamics of groups composed solely of humans, followed by modelling human intuitions and reasoning, such that the latter model can be used for simulating human-agent hybrid systems. In this work, as a first step towards understanding a hybrid system, we develop a cooperative game that is played by human subjects on a virtual network. First, we observe how large groups can emerge, using the concept of a connected cluster of nodes. Then we construct a probabilistic model that captures the human decision making pertinent to the game, and simultaneously allows us to analyse the bounds on rationality and cognition of the players.  

We choose a framework that is in the spirit of the earlier works by Kearns {\it et al.} (see \cite{kearns2012behavioral} and references there in). In a series of experiments they studied the effect of network structure on the efficiency of solving problems like the graph coloring and consensus by human subjects \cite{kearns2012behavioral,kearns2006experimental,judd2010behavioral}. They also showed that the amount and quality of information in the system that is available to the subjects influences their performance, depending on the structure of the network over which the subjects are interacting. Our game also has similarities to the extensively studied matching problem that considers two distinct sets of individuals, like, men-women, producers-consumers, and employers-job seekers, from which members get matched in pairs to their own mutual benefit \cite{gale1962college,laureti2003matching}. The latter problem has also been studied by Coviello {\it et al.}   \cite{coviello2012human} in the form of a distributed coordination game on a network where players get rewarded when all the nodes get matched in pairs. They found that human subjects participating in the experiment tend to behave prudently while averting risks, and this behavioural trait influences strongly their performance and capabilities to complete the expected goal. The complex relation between network properties, human behaviour and collective performance in problem-solving tasks has been explored in different works \cite{baronchelli2010modeling,guazzini2015modeling, centola2015spontaneous}, and it has been shown that coordination, cooperation, and other social interactions within human groups can be described and analyzed through carefully designed experiments involving human subjects. In the current set-up the players are incentivized to arrange themselves in groups rather than pairs. Closely knit groups or communities are ubiquitous in the society that act as chambers of collaboration and innovation across diverse fields of human endeavour like performing arts, science, and technology development  \cite{girvan2002community,wuchty2007increasing,lambiotte2009communities,lakhani2010topcoder}. Therefore, the game can be also placed in the context of game-theoretic studies of social group formation, for example, games \cite{chauhan2018Schelling,elkind2019schelling} that are based on Schelling's segregation model \cite{schelling1971dynamic}, and more generally, hedonic coalition formation games \cite{dreze1980hedonic,aziz2016hedonic}.   

During the rounds of the game the players coordinate their actions in pairs by communicating over the network links, and exchange their locations or node positions. We focus on the problem of complex decision making by human players linked with others in a network, where the players have information at the local and global scales, i.e. colours of their neighbours and cluster sizes, respectively. The model is able to explain the decision making in terms of the perception of risk by the players, as well as the cognition of their neighbourhood information. The model also provides insight into possible situations where excessive caution by the players would hinder their mobility and aggregation behaviour aimed towards group formation. On the other hand the lack of prudence or cautiousness would allow the groups getting fragmented. Finally, we use our model to scan the parameter space and compare other possible strategies with the strategy employed by humans.

\begin{figure*}[t]
\centering
\includegraphics[width=1.0\textwidth]{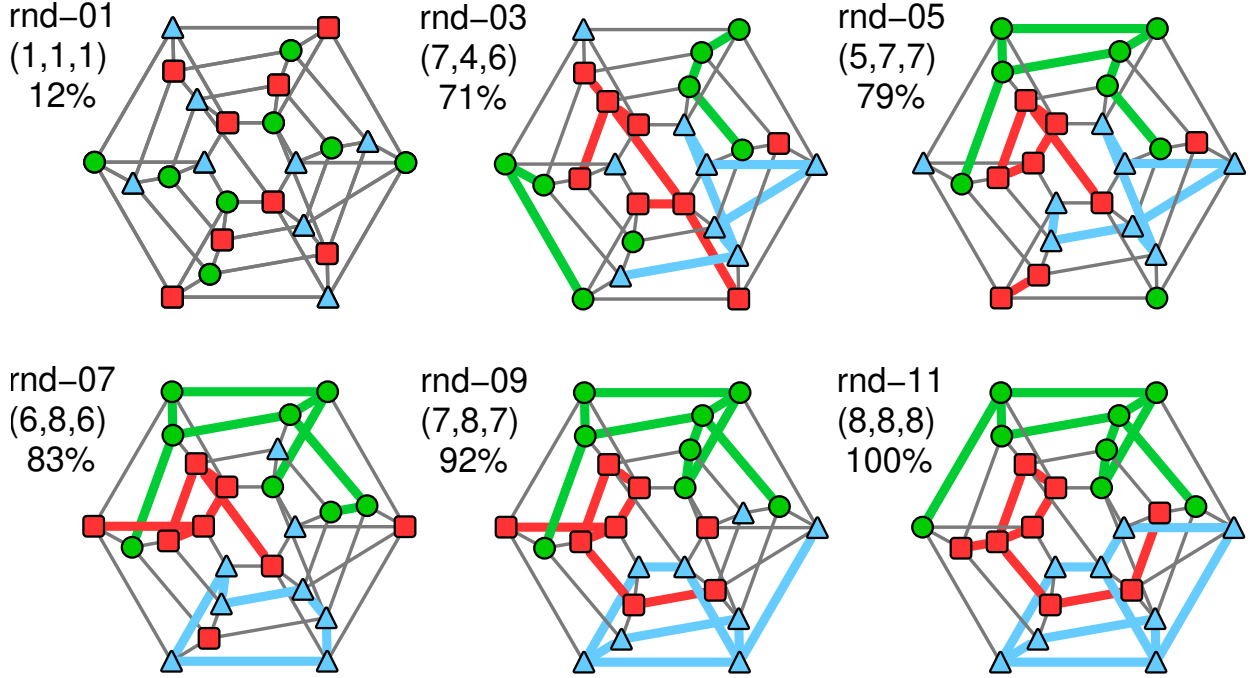}
\caption{{\bf A game with 24 players.} Starting from a graph-colored configuration, neighboring players exchanged places over 11 rounds, after which a solution is achieved, i.e. after the 11th round the three largest clusters belonging to each colour have reached the maximum size of 8. Within a given cluster (for a given colour) any pair of nodes will have a path exclusively through the nodes belonging to the cluster. The colours attributed to the subjects are shown in the figure. In addition, links that connect the players in a given cluster are marked with the corresponding colour while the other links are shown in gray. Only rounds with odd numbers are shown and denoted by the label `rnd-'. The largest cluster size for each colour at the end of the corresponding round is indicated as a triplet $(S_r$,$S_g$,$S_b)$ with $S_r$,  $S_g$, and $S_b$ being the largest cluster size corresponding to red, green, and blue, respectively. The network consisted of a regular squared mesh of $6\times 4 = 24$ nodes, with periodic boundaries and three small-world links. The average of the largest cluster sizes of the three colours is scaled by the maximum value (8) and expressed as a percentage is also indicated for each round.}
\label{fig1}
\end{figure*}

\section{Methods}
\label{Methods}
\subsection{Game rules} Human subjects playing the game are grouped into $m$ classes having $l$ individuals in each class, such that in sum there are $n = ml$ individuals. A class is denoted by a colour, e.g., red, blue, green, etc. The game is played on a connected network of $n$ nodes. Each one of the $n$ nodes of the network is occupied by only one player, hence there are no empty nodes in the network.  The goal of the experiment requires that by swapping places during successive rounds, players end positioned on the network with $m$ clusters each of size $l$. A connected red cluster implies that from any red player any other red player can be reached by a connected near-neighbour path consisting of only red players (see the last network in Fig. \ref{fig1}). 

The game is played in rounds, and in each round players decide on exchanging their locations with other players in the network, in a requesting-accepting process. In each round, one of the $m$ colours is chosen, and players having this colour act as requesters. Players having other colours receive the possible requests. Each round consists of two distinct stages:  a first stage with requests being sent by players of the chosen colour, who want to move from their current locations. A requester can send a request to only one of the neighbouring players who has a colour different from its own (therefore, players of the same colour can not exchange places among themselves). In the second stage, acceptances occur, where those players who received requests in the previous stage from the requesters, decide on whether to accept or not, one of the received requests for exchanging places. If a request is accepted, those players involved (i.e. requester and acceptor) exchange their locations in the network. After all the possible exchanges are made, the round ends. In the next round a different colour is chosen
for the requesting activity. The colour chosen for the requesting activity on each round changes in a cyclic way, such that in $m$ rounds, each of the $m$ colours have been chosen once. The rounds continue till the goal is achieved or the number of rounds reaches the maximum limit, whichever happens earlier.

\begin{figure*}[t]
\centering
\includegraphics[width=0.95\linewidth]{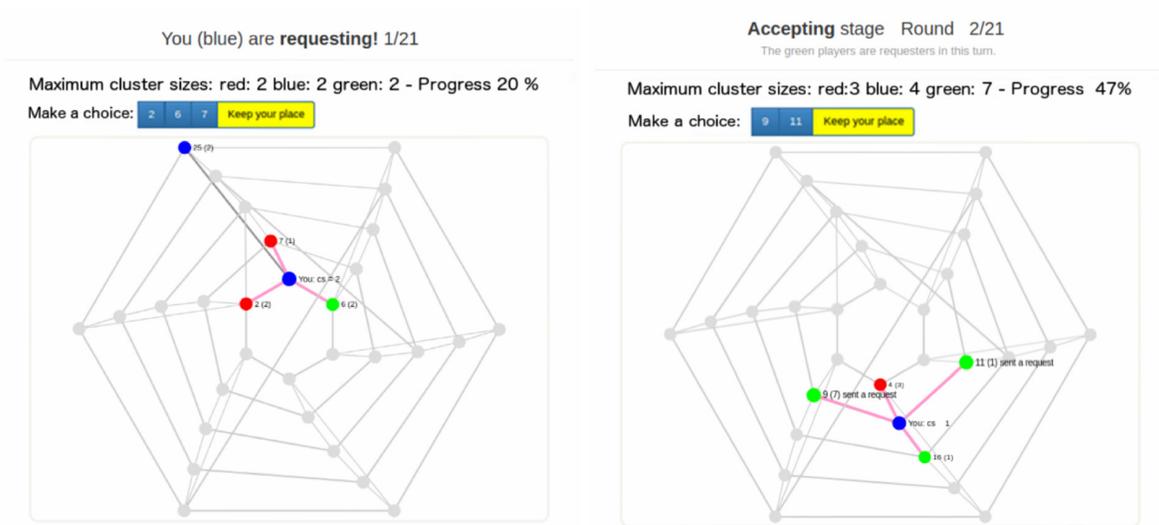}
\caption{{\bf The graphical user interface for the game during requesting (left) and accepting (right) stages.} The focal subject (``YOU'') is able to view the entire network but the colour of nodes that are not immediate neighbors are masked. Links to the neighbors that have colours different from the focal subject are shown in magenta and links to subjects having the same colour are shown in dark grey. Other links are shown in grey. Buttons allowing to choose between options are displayed above the network. Adjacent to the neighbouring nodes the following two numbers were provided -- (i) an identifier for the node location, and (ii) the cluster size of the neighboring player located at that node (within brackets). 
}
\label{fig2}
\end{figure*}

During each round, a player is provided the following information:

\begin{enumerate}
\item The colour of other players in the neighbourhood.
\item The size of the player's own cluster.
\item The size of the cluster of each of the neighbouring players.
\item The largest cluster size for each of the $m$ colours.
\item The average collective progress ($ACP$) defined as the average of the sizes of the largest  cluster in each colour normalized by the maximum possible size $l$. In case $m$ largest clusters reach sizes $l$ the $ACP$ is $1.0$ (desired outcome).
\end{enumerate}
The game is terminated once the desired outcome is reached, i.e $m$ clusters of size $l$ are generated, or after a fix number of rounds $T$, where $T$ is a multiple of $m$, which allows all the colours to have the same number of rounds to be requesters. Incentive per game is based on a score that is calculated from the $ACP$ (see the following section).

\subsection{Experimental setting}

The experiment was conducted in a single session, in a computer lab at the Department of Computer Science of Aalto University located in Espoo, Finland, on the 9th of August, 2017 with 30 individuals recruited from an online volunteer pool and from advertisements in social media. Before the experiment all the subjects were provided with information sheets as well as sheets for informed consent. The informed consents, once signed by the subjects, were collected before the experiment began. No personal information of the subjects was collected other than contact emails for rewarding purposes. To help the participants  understand better the game dynamics and its rules, a short explanatory visual presentation was given in the room to all the groups and, before the sessions started, each participant played a trial game lasting 6 rounds to ensure that the subjects got acquainted with the user interface of the game. During the sessions the presentation file was accessible to the subjects on their desktop. The file is available in the supplementary material (SM).
The interface and engine of this online game was implemented using the oTree framework (Fig.~\ref{fig2}) \cite{otree}. Subjects' view of other workstations was restricted and all communication during the games was forbidden.  

During the four-hour session, a total of 9 games were played, consisting of five games with 24 players and four games with 30 players. We ensured that during each of the first 5 games (24-players game), six of the thirty participants would skip exactly one game. For each game, the network consisted of a regular squared lattice ($4\times 6$ or $5\times 6$) with periodic boundary conditions. We had three additional long-ranged links (not resulting in triangulations) to introduce a small-worldness. In every game the positioning of these links were altered. Also the places of subjects at workstations were shuffled between games. The networks used in the game can be considered as realizations of the Kleinberg model \cite{kleinberg2000small} in the limit that small-world links appear independent of distance. The time given to the players to make a decision (either requesting or accepting)  was changed with rounds progressing, being 30 seconds during the first 5 rounds and 20 seconds during the rest.

\begin{figure*}[t]
\centering
\includegraphics[width=0.95\linewidth]{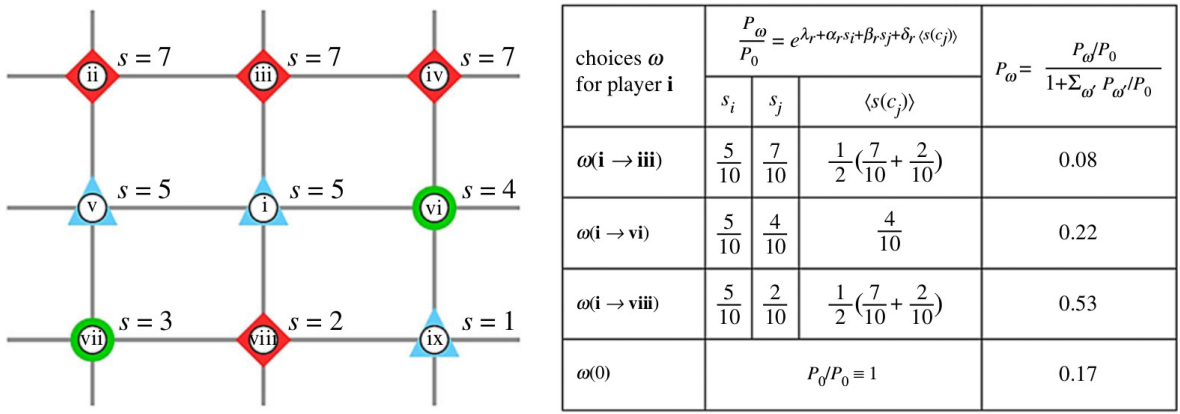}
\caption{Example showing how the model assigns probabilities to an agent in the requesting mode in a given neighbourhood. (Left) Hypothetical instance of a neighbourhood around a focal agent with label {\bf i}. Agent {\bf i}'s color is blue. Agents {\bf iii}, {\bf v}, {\bf vi} and {\bf viii} are immediate neighbors of agent {\bf i}. The sizes of the cluster to which each agent is connected is shown to the right of each agent. Note that agent {\bf i} has 2 neighbors of color red (agents {\bf iii} and {\bf viii}), a neighbor of color green (agent {\bf vi}), and one neighbor of the same color (agent {\bf v}). (Right) Probabilities calculated using the model according to which {\bf i} selects  one of the 4 possible choices ($\omega$), namely to send a request to one of the agents {\bf iii}, {\bf vi} and {\bf viii},  or to not send any request and retain its place in the next round.The last option is denoted by  $\omega(0)$, and represented by $P_0$ in the model. The parameter values $\alpha_r$, $\beta_r$, $\delta_r$, and $\lambda_r$ are taken from Table~\ref{tab:param}.}
\label{fig3}
\end{figure*} 

The incentive offered for participating in the experiment were movie tickets, sent to the players by email after the whole session was completed. For a given network topology and initial positioning of subjects we also calculated the quantity $ACP_{rand}$ from simulations where agents play randomly. We obtained a normalized score for each game by scaling $ACP$ with respect to $ACP_{rand}$ and accumulated the total score ($S$) over 9 games as   
\begin{equation*}
S = \sum_{k=1}^9 \frac{ACP(k)-ACP_{rand}(k)}{1-ACP_{rand}(k)}.
\end{equation*}
To each subject one movie ticket was given for participation and $4\times S$ (nearest integer) movie tickets were given as reward.

\subsection{Model}
\label{Model}
During the game the players take decisions on sending (or not sending) requests and accepting (or not accepting) requests based on the available information of their immediate neighbourhood, and the sizes of the largest clusters of each colour. The decision making logic employed by the agents is expected to be heterogeneous yet far from random. It is known that exact coordination between two agents is possible in the presence of common knowledge \cite{halpern1990knowledge}. However, in this game the knowledge of a subject and any of its neighbours do not completely overlap.  We construct a model that is based on the notion of probability matching \cite{vulkan2000economist,shanks2002re,arganda2012common,vul2014one,krafft2016modeling}. This model is expected to serve both the purposes of uncovering the decision making logic of the subjects as well as act as template for agent based simulations.

For an agent $i$ having a colour $c_i$ we assume that its neighbourhood $\mathcal{S}$ can be uniquely characterized by the following quantities -- (i) the current cluster size $s_i$, (ii) the set of cluster sizes $\{ s_j \}$, where $j$ is a neighbour of $i$ having a colour different from $i$, that is, $c_j\ne c_i$. We place all the neighbourhoods $\mathcal{S}$ that have identical supersets $\{s_i,\{ s_j\} \}$ (or that can be made identical by the ordering of $j$'s) into a given category $\mathcal{C}$. 

In a given category $\mathcal{C}$ we consider an agent $i$ with the set of neighbours $\{j_1,j_2,j_3,...\}$ that have colours different from $i$. In the requesting phase it has to choose from the following set of options $\omega$: $\{$stay at current location without sending a request, send a request to $j_1$, send a request to $j_2, ...\}$. We assume that with each option $\omega$ the player associates a probability, $P_\omega$ of it being beneficial to the progress of the game. Here beneficial may refer to an increase in the cluster size for the colour of $i$, or increase in the cluster size for colour of $j$, or both. Therefore, using probability matching, the probability of choosing an option $\omega$ is given by, $p_\omega=P_\omega/\sum_{\omega'} P_{\omega'}$. Restricting $\omega$ to the set of options when a request is sent one can write, 
\begin{displaymath}
p_\omega=\frac{P_\omega/P_0}{1+\sum_{\omega'}P_{\omega'} /P_0},
\end{displaymath}
where $P_0$ is the estimated probability that not sending a request (that is, not moving) is a beneficial option.

Next, we focus on different categories that could be realized during the course of the game, and for each category we identify the cases (characterized by neighbours with cluster size $s_j$) when one or more requests were sent from the focal players $i$ to players $j$. These sets are indeed options that were actually executed. Thus in each category and for each option $\omega(j)$ we accumulate the total number of cases ($N_\omega$) when a request was sent. Additionally, in the same category we accumulate the total number of cases ($N_0$) when no request was sent, and calculate the ratio $(N_\omega/N_0)/(1+N_\omega/N_0)$, where this ratio lies in $\left[0,1\right]$. Taking this ratio as the dependent variable we perform a logistic regression on the following set of variables, (i) $s_i$, (ii) $s_j$, and (iii) $U_j$ with data from all chosen $\omega$'s from all categories. The quantity $U_j= \langle s(c_j) \rangle-s_j$, where $\langle s(c_j) \rangle=\sum_{k=1,n_i(c_j)} s_k/ n_i(c_j)$ is the average of the cluster sizes of the $n_i(c_j)$ neighbours of $i$ that have the same colour as $j$. We use this difference $U_j$ as the measure of disparity that a focal player could possible recognize in the neighbourhood in the requesting mode. A large value of this difference might encourage the focal player to send a request to a player having colour of $j$, so that clusters having colour $c_j$ could merge. There could be better measures of disparity but our choice is guided by the linearity of the model. For the actual fitting instead of $U_j$ we use $\langle s(c_j) \rangle$ as $s_j$ is already an independent variable. Although we base our model of decision making on probability matching, it could as well be considered as a log-linear response model \cite{blume2001interactions,young2015evolution}. The fit to a logistic function allows us to use the following expression for a requester, $P_\omega / P_0=\exp\{\lambda_r+\alpha_r s_i+\beta_r s_j+\delta_r\langle s(c_j) \rangle\}$ from the estimation $P_\omega / P_0= p_\omega / p_0 = N_\omega / N_0$, where,  $\alpha_r$, $\beta_r$, $\delta_r$, and $\lambda_r$ are parameters corresponding to a requester. Once evaluated by fitting to the data from the experiment, these parameters are used in numerical simulations of agents as shown in Fig.~\ref{fig3}. A similar scheme is used for data from acceptors, and the corresponding parameters $\alpha_a$, $\beta_a$, $\delta_a$, and $\lambda_a$ are evaluated.


\section{Results}
\label{Results}
\subsection{Experiment} 
\label{Experiment}

For our experiment we chose $m$ = 3, i.e. colours red, green and blue. We conducted 5 games with $l$ = 8 ($n = 24$) and 4 games with $l$ = 10 ($n = 30$) with $T=21$. The network was taken as a square lattice with periodic boundary conditions and additional small world links. For the initial condition, we tried to position the players in a graph coloured configuration, that is, a subject with a given colour (say, red) is surrounded only by neighbours who have a different colour (i.e. can not be red). All the games with 24 players and three out of the four games with 30 players reached the desired outcome within 21 rounds. The evolution of the clusters in one of the games with 24 players is shown in  Fig.~\ref{fig1}. This particular game finished in 11 rounds. The games with 24 players took $9$ rounds to complete on average, and the completed games with 30 players took $16$ rounds on average. 

\begin{figure*}
\centering
\includegraphics[width=0.95\linewidth]{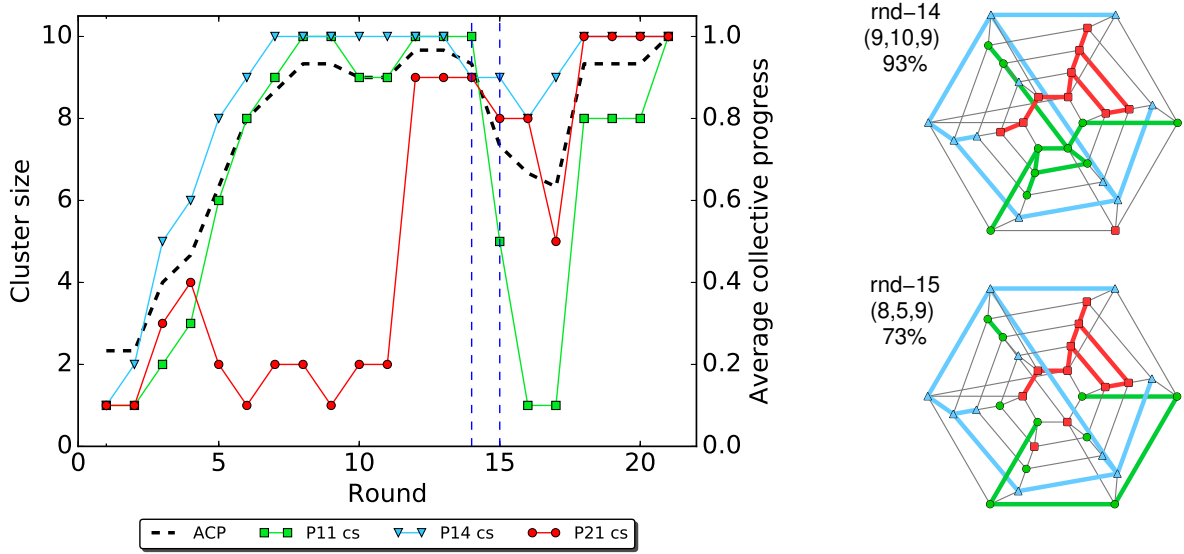}
\caption{\label{fig4}
{\bf A typical time evolution of cluster sizes in a game with 30 players.} Cluster sizes of three randomly chosen players corresponding to the three colours (red - circles - player 21, green - squares - player 11 and blue - triangles - player 14) are shown as the game progresses from round 1 to round 21 (final round). The average collective progress (normalized value of the average of the largest cluster sizes of the three colours, ACP; see Methods) is denoted by the dashed line. The typical dynamics leading to a solution of the problem can be observed from the evolution of the network, with an initial stage of fast progress  ($\approx0.9$ after 8 rounds), followed by a stagnation regime where only a small number of location exchanges happen in the network, eventually leading to the formation of the required three maximum-sized clusters. In the stagnation regime players that are trapped in smaller clusters are facilitated to move such that they eventually merge with larger clusters. For this particular game, we illustrate such an event occurring at round 14. The configurations during rounds 14 and 15 are shown to the right with legends similar to those used in Fig.~\ref{fig1}. A player with a red colour and having cluster size of 1 (located at the bottom right corner of the mesh at round 14) exchanges place with another player with colour green. As a result of which the green cluster is fragmented. We observe this as the (chosen) green player's cluster size decreases from the maximum possible size (10) at round 14 to the minimum possible size (1) in the next two rounds. However, the exchanges occurring after the 15th round allows the players reorganize rather quickly to the desired configuration.}
\end{figure*}

\begin{figure*}[] 
\includegraphics[width=0.9\linewidth]{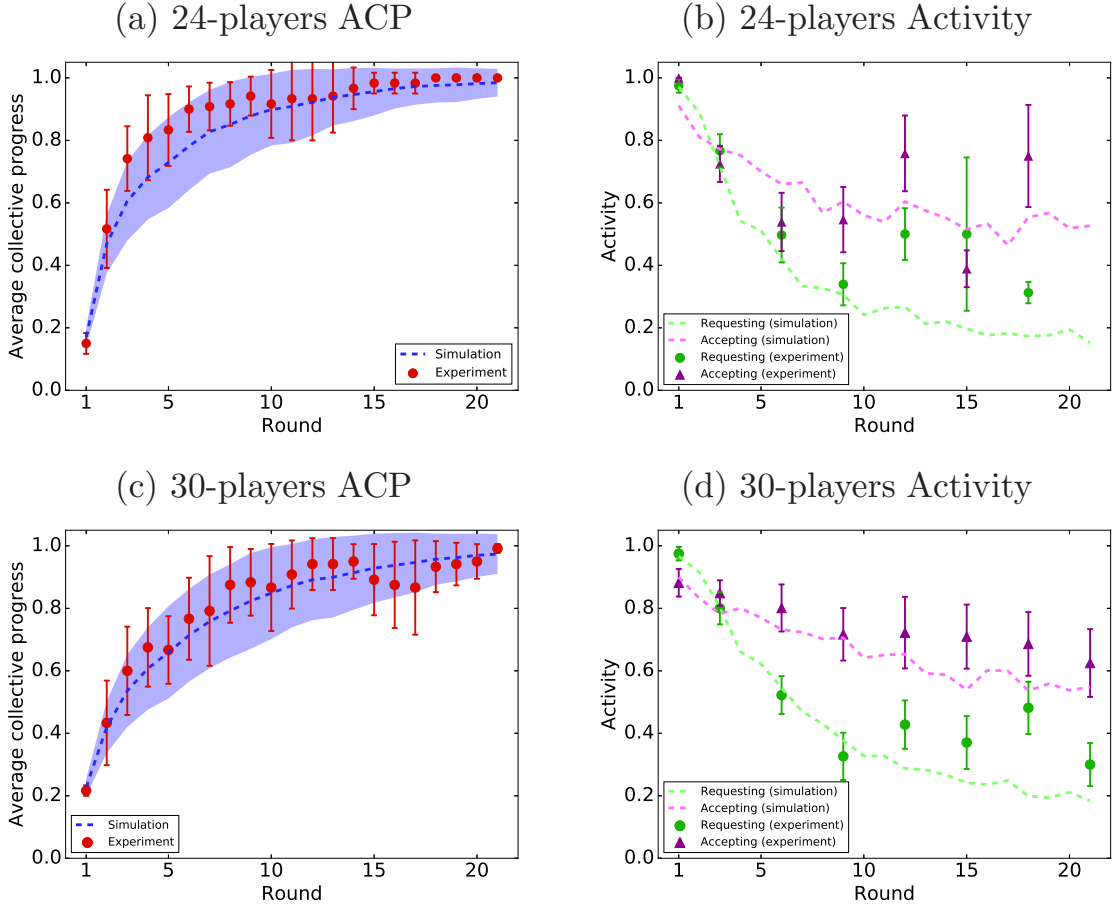} 
  \caption{Comparison between experiments and simulations using the model for games with 24 and 30 players. (a and c) The average collective progress (normalized value of average of largest cluster sizes corresponding to the three different colours) is plotted against the round number. The points are binned values from the 5 experiments in the case of 24 players and from the 4 experiments in the case of 30 players. The dashed lines are the result from simulations of the model with 500 runs. (b and d) The requesting and accepting activities in the games. The requesting activity is measured as the the ratio between number of actual requests and the maximum possible requests per round (circles). Similarly, the accepting activity is measured as the ratio between the number of accepted requests and the total number of requests received (triangles). The dashed lines are results from simulations. The error bars in the figures indicate the standard deviations.}
  \label{fig5} 
\end{figure*}

In general, all the games showed a fast initial growth in the sizes of the largest clusters. After which the overall activity in terms of requesting and accepting decreased. In this phase the players appeared to become ``conscious'' of the presence of players in the vicinity who were trapped in smaller clusters. This resulted most likely from the understanding of the collective goal of the experiment. As a result players in larger clusters cooperated with the isolated players and with players in small clusters by exchanging locations, which sometimes caused fragmentation of the larger clusters. Such a case is shown in Fig.~\ref{fig4}. The $ACP$ and the activity for the games are shown in Fig.~\ref{fig5}. It is observed that the clusters become less and less active as the game progresses in the wake of more passive requesting and accepting. 


\subsection{Numerical results}
 The fitting parameters are provided in table~\ref{tab:param}. All the parameters turn out to be significantly different from zero with $p<0.01$. As we have two sizes for the networks, for the purpose of fitting we use cluster sizes that are scaled by the maximum cluster size, in such a way that the cluster size variable varies between 0 and 1. As can be observed from table \ref{tab:param}, the corresponding parameter values for the two sets (i.e. with 24 players and 30 players) are mostly within the error bars. Therefore, using the data on all the 9 games we recalculate the coefficients. Using these coefficients we simulate the model (see Fig.~\ref{fig3}). With 24 agents and 500 simulations around $82\%$ of the games reach a solution within 21 rounds; and the games end with a mean $ACP$ of $0.98$. In case of 30 agents $66\%$ of the games reach a solution and the mean $ACP$ is $0.95$. In Fig.~\ref{fig5} we compare the $ACP$ and the activity from the experiment with those from the simulations (averaged over runs) using the model. Note, that in addition to the probabilistic choice  we use an additional rule that prevents exchanges between agents belonging to different large clusters and increases the fraction of games completed. This rule, however, has marginal effect on the $ACP$ and the overall activity (see the following section). 

\begin{figure*}[h]
\centering
 \includegraphics[width=0.9\linewidth]{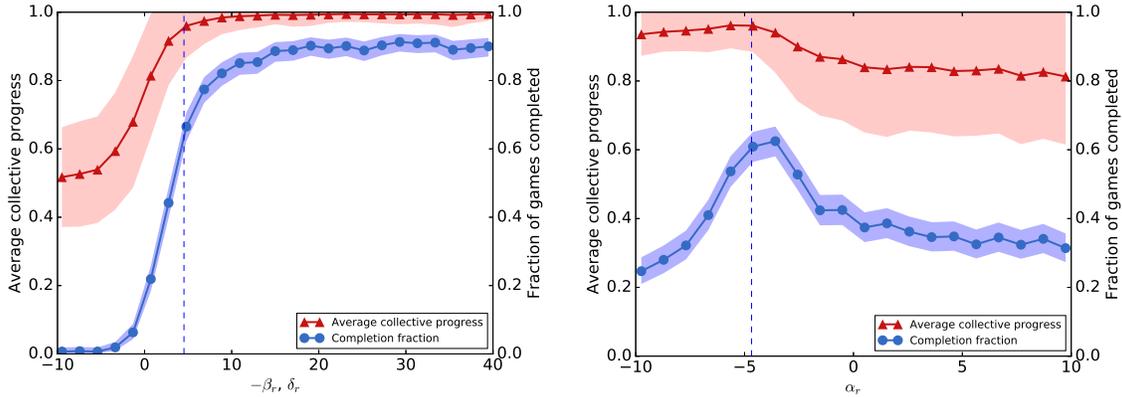}
\caption{{\bf Sensitivity of the requester model.} Results from simulations of games with 30 agents demonstrating the effect of variation of the parameters in the model corresponding to the requesters on the average collective progress (red triangles) and the fraction of games completed (blue circles). (Left) The parameters $-\beta_r$ and $\delta_r$ are assumed to be equal and simultaneously varied. The dashed vertical line shows the location of $-\beta_r$ obtained from the experiment. (Right) The parameter $\alpha_r$ is varied. The dashed vertical line shows the location of $\alpha_r$ obtained from the experiment. Each of the points in the figures are averaged values from 500 runs. The shaded region in case of the ACP represents the standard deviation associated with the points. In case of the completion fraction the region indicates the Clopper-Pearson interval.}
\label{fig6}
\end{figure*}

Interestingly, the fact that for requesters and acceptors the magnitudes of $\beta$ and $\delta$ also overlap within the error margins, could support our initial ansatz about the inclusion of the term $U_j=\langle s(c_j) \rangle-s_j$, as an independent variable, which would, however, diminish the importance of $s_j$ as a separate independent variable. To investigate the importance of such a term, we slightly modify our model by considering the magnitudes of $\delta_r$ and $\beta_r$ to be equal. We simulate the model by varying  $\delta_r$  (keeping the relation $|\delta_r|=|\beta_r|$) while taking the values for the other coefficients from the table~\ref{tab:param}. The resulting plot is shown in Fig.~\ref{fig6} (left). The plot shows that a larger $\delta_r$ (and $\beta_r$) enhances the performance of the agents. Similarly, by varying $\alpha_r$ in the model we benchmark the perception of risk in the human subjects. Agents in the model are less likely to breakaway from clusters when $\alpha_r$ is negative and large. The Fig.~\ref{fig6} (right) shows a region near  $\alpha_r=-5$ where the agents perform best. Remarkably, we find that $\alpha_r$ obtained from the experiment coincides with the optimal value.

\subsection{Model details}
\subsubsection{Stability rule for clusters}
\label{model-details-stability}

\begin{figure*}[h]
\centering
 \includegraphics[width=0.8\linewidth]{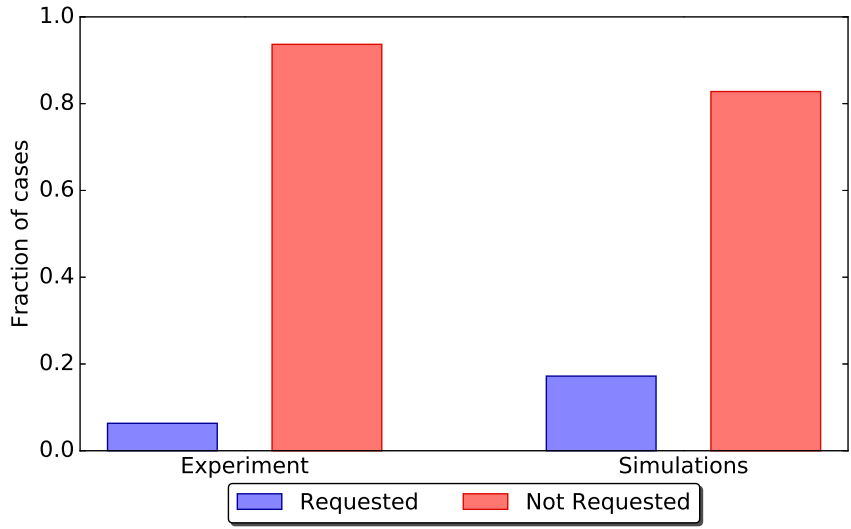}
\caption{Requesting between large clusters in the games with 30 players -- experiment (columns on the left) and simulations without the stability rule (columns on the right). Requesters having cluster sizes larger than 6 are considered (in this case the largest possible cluster size is 10). The cases when such requesters sent requests to other players having cluster size larger than 6 are counted. Also the cases when such requesters, though positioned in the neighbourhood of players with cluster sizes larger than 6, did not sent any request to the latter are counted. The above two counts are normalized by the total number of cases and compared. The plots reveal that when simulating with the basic model, there is an excess of cases where players belonging to larger clusters send request to players in larger clusters.}
\label{fig7}
\end{figure*}

The expressions for probabilities in our basic model is a continuous function of the variables and is linear in terms of the argument. Actual human decision making can be quite complex and more accurate description might require inclusion of non-linearities or discontinuous dependence in terms of the variables. A lack of this in the basic model might have resulted in the excess occurrence of requesting activity between large cluster sizes. In Fig.~\ref{fig7} we show this by comparing experiment and simulations for the games with 30 players. We observed that in the model the presence of such activity can impact the completion of the games. Requesting (and accepting) activity between players belonging to large clusters could be considered as detrimental if not useless in terms of reaching the solution. In case the large clusters become unstable, the solution might not be reached. Therefore, in the simulations we prevent such requesting actions from taking place. We prohibit any requesting activity between two agents having cluster size larger than $0.6l$, where $l$ is the maximum possible value for a cluster size. The effect of such a rule can be investigated by introducing a parameter $f$ in the model such that $1-f$ is the probability of allowing such an action. With $f=0$ we have the basic model and with $f=1$ such requests are completely forbidden. In Fig.~S1 (SM) we show the effect of varying $f$ on the $ACP$ and the fraction of games completed. For the results of simulations reported in the main text we take $f=1$. Note, that large clusters can still fragment when requests come from smaller clusters.   

\subsubsection{Acceptance in the model}
\label{model-details-acceptance}

We observed that in the experiment the probability for a player to receive more than 2 requests in a given round was negligible. The dominating case was that of a single request ($70\%$) as can be observed from Fig.~S2 (SM). Also, we found that when simultaneously two requests were received, the choice of acceptance by the focal player was not significantly influenced by the cluster sizes of the requesters (Fig.~S2, right). Therefore, we modeled acceptance by fitting our basic model (having the same structure as that of requesting) to the dominating case of single requests. In the simulations when there were cases of multiple requests we randomly considered one of the requests to be weighed against the decision of not exchanging.  

 



\begin{table*}[h]
 \caption{Values of parameters (coefficients and the intercept) derived from a logistic regression on the data.}
 \centering
 \resizebox{\textwidth}{!}{%
 \begin{tabular}{|@{\hskip 0.1in}c@{\hskip 0.1in}|@{\hskip 0.1in}l@{\hskip 0.1in}|@{\hskip 0.1in}c@{\hskip 0.1in}|@{\hskip 0.1in}c@{\hskip 0.1in}|@{\hskip 0.1in}c@{\hskip 0.1in}|@{\hskip 0.1in}c@{\hskip 0.1in}|}
 \hline
 Focal & Variable  & Coeff & 24-player & 30-player & Joined\\
 \hline
 \hline
 \parbox[t]{7mm}{\multirow{4}{*}{\rotatebox[origin=c]{90}{Requester}}} & Cluster size of requester  & $\alpha_r$ & $-5.33 \pm 0.71$ & $-4.06 \pm 0.52$ & $-4.68\pm 0.41$ \\

& Cluster size of requested neighbour  & $\beta_r$ & $-5.36\pm 1.58$ & $-3.80 \pm 1.01$ & $-4.52\pm 0.84$\\

& Average cluster size (neighbour's colour)  & $\delta_r$ & $4.89 \pm 1.89$ & $2.91 \pm 1.19$ &$3.85\pm 0.99$  \\

& (Intercept) & $\lambda_r$ & $3.25\pm 0.59$ & $2.61\pm 0.43$ & $2.96\pm 0.34$ \\
 \hline
  \hline
 \parbox[t]{7mm}{\multirow{4}{*}{\rotatebox[origin=c]{90}{Acceptor}}} & Cluster size of acceptor  & $\alpha_a$ & $- 5.23 \pm 0.83$& 	$- 3.64\pm 0.54$  & $-4.15 \pm 0.45$ \\

& Cluster size of requesting neighbor  & $\beta_a$ & $- 5.14\pm 2.03$ &	$- 4.53 \pm	  2.17 $ & $-4.39\pm 1.50$ \\

& Average cluster size (neighbour's colour)  & $\delta_a$ &   $5.32 \pm	  2.33$ &	 $ 5.19 	\pm  2.30$  & $4.76\pm 1.60$ \\

& (Intercept) & $\lambda_a$ & $3.25 \pm 0.62$ & $2.31 \pm 0.40$ & $2.70\pm 0.33$ \\ 
 \hline
 \end{tabular}
 }
 \label{tab:param}
 \end{table*}

\section{Discussion}

Models and experiments based on game theory \cite{doebeli2005models,nowak2006five,galeotti2010network}, especially those based on Prisoner's Dilemma (PD) have been used extensively to study the formation of groups \cite{helbing2009outbreak,suri2011cooperation,wang2012cooperation,antonioni2015short}. In this paradigm the existence of cooperative ties could lead to the formation of cohesive groups, and the risk arises when players choose to defect. However, in the present framework we interpret risk as an individual player's decision to favour exchanging their current locations. Unlike models based on PD where the payoff matrix would ideally determine the amount of risk, in our case it is the propensity of an individual to relocate depending on his or her current cluster size. The dilemma arises from incomplete information, but the goal is always collective, which is how to converge to a win-win situation in a limited number of rounds.

The results from the experiment show, in general, that human players are able to process complex information about their neighbourhood as well as take into account global information about cluster sizes. It appears that that the players understand well the cooperative setting of the game. In order to achieve the desired target the players tend to coordinate their actions such that an exchange is done if it is mutually beneficial. The limited capacity of processing the neighbourhood information is apparent in Fig.~\ref{fig6} (left). It is found that larger magnitudes of the requester parameters of $\beta_r$ and $\delta_r$ would have signified a `wiser' decision making and would have lead to faster progress in the game and better completion rates. However, the fact that the values of the parameters (see table~\ref{tab:param}) are significant with appropriate signs,  reflects the rationality in the choice of neighbours during the action of request. The values of the acceptance parameters $\beta_a$ and $\delta_a$ would also signify rationality although most of the times the acceptors receive only one request and the decision is limited to agreeing or disagreeing to exchange places. 


The dynamics of fragmentation and formation of clusters crucially depends on $\alpha$ (the coefficient of cluster size of the focal player, the latter being a requester or an acceptor). Whereas a negative value of $\alpha$ would make larger clusters stable, an extremely large magnitude would make clusters inactive and render exchanges impossible. Interestingly, Fig.~\ref{fig6} (right) shows that for requesters the value of $\alpha_r$ obtained from the experiment coincides with the optimal value predicted from simulations using the model. This kind of balance could result from the cooperative setting, which was well comprehended by the human subjects. This is evidenced from the cluster fragmentation illustrated in the examples shown in Fig.~\ref{fig4}. Variation in $\alpha_a$ does not seem to have much effect. Overall, the parameter values could be reflecting as strategies of human individuals for engaging in coordination when risks are present \cite{thomas2014psychology}. We have also examined the aspect of learning during sessions. To do this we partitioned the decisions into categories and checked for significant differences between their frequency distributions across the 9 sessions employing chi-squared tests. The differences appeared to be non-significant (see the SM, smallest p-value is 0.07). Similarly, we have extended the basic model to quantify the random effects resulting from the heterogeneity between the participants. By using mixed effects logistic regressions we obtained standard deviations corresponding to the parameters of the original model (also provided in the SM).

 The choice of the network was primarily guided by the issue of achieving convergence in a limited number of rounds while maintaining a certain level of complexity. From each player's point of view, the accessible information comes from two sources: the local neighborhood which describes the status of his/her neighbours, and the global information, which only shows the overall progress towards the goal, without giving any hint about the configuration of whole the network. Preliminary agent-based simulations on other topologies, for example networks with structural communities \cite{lancichinetti2008benchmark}, revealed extra complications and bottlenecks for the players. In general, non-regular networks with varying neighborhoods are expected to increase the difficulty for players trying to locate themselves and their neighbors in the network. Also, frequent changes in the number of neighbors after each move would add up to the mental processing required in each round. In this regard, the chosen small-world network could have also taken different parameters (for example using as an initial backbone a ring of size $3 \times 10$ instead of the chosen $6 \times 5$). Simulations allowed us to assess the possible layouts and possible number of long distance links that could facilitate solving the problem in a limited amount of time. The choice of three colors was, again, with the idea of keeping the game complex enough without being extra demanding. We ran simulations of the game with more colours, and it always required considerably more rounds to be completed. Besides, increasing the number of colors would have increased the number of variables to deal with for each player, requiring more time for processing information and make decisions. 

We place the game in the broad context of group formation in social networks and technological networks. The experiment illustrates how individuals located on a network and coordinating over links could achieve configurations that would in principle benefit all. In the context of real-world social networks such dynamics would represent mutual enhancement of individuals by the `bonding' and `bridging' of social capital \cite{granovetter1977strength,larsen2004bonding,ellison2007benefits} or the formation of social coalitions \cite{dreze1980hedonic,aziz2016hedonic}. 
In a sense, the notion of a link between two nodes having the same colour is comparable to the notion of an edge in a puzzle graph (representing a compatible idea) in the recent formulation of `jigsaw percolation' \cite{brummitt2015jigsaw}. While an exchange of positions between two individuals is to be understood primarily as simultaneous changes in the social space of the pair, such an activity could also be considered as an exchange of physical locations of two individuals like that in a faculty exchange program between universities. The desired nature of linking in the network with its underlying spatial structure can also be considered relevant in communication networks where the nodes are autonomous mobile agents establishing peer-to-peer radio network. The theory of cooperative games has been used to design deployment protocols of mobile agents where a coalition of agents would share a certain frequency spectrum \cite{saad2009coalitional}.

As we have performed the experiment with a limited number of subjects, the behaviour captured via the coefficients in the model will reflect their particular characteristics and may not be universal. For a different subject pool the individual coefficients may be different while the overall behaviour may still be close to optimal. However, we have no reason to suppose that the modelling scheme, maintaining the current level of complexity, would be entirely different and that the current formulation would exert any special influence on our inferences. Overall, we expect the broad principles revealed in this study to apply more or less universally. Our decision making model not only serves the purpose of extracting the behavioural aspects of real human subjects, it allows us to compare other possibilities with respect to the parameter values. Although we do not perform an exhaustive search in the parameter space, we gain sufficient insight into human behaviour when we find faster convergence for some parameters. Our goal in the near future is to perform experiments with artificial agents or bots playing with humans \cite{shirado2017locally,crandall2018cooperating}. Agents formulated using the model can mirror the typicalities of human behaviour who are actually willing to cooperate during tasks that require collective coordination \cite{jennings2014human}. Such agents may be used to test the collective performance in human-agent hybrid systems where humans are guided to make decisions based on less collective or selfish motivations. For instance, it would be interesting to compare the outcomes from games where the individual groups are entirely constituted by humans or bots, with games where groups are formed by mixing humans and bots.

\section*{Data accessibility}
All the relevant data behind this paper has been deposited in a public data depository: \url{https://doi.org/10.5281/zenodo.3237575}. The codes for the simulation is available at \url{https://github.com/ttakko/NetworkGameSimulations}.

\section*{Ethics permission}
Approval for the experiment was obtained from the Aalto University Research Ethics Committee. The experiment was performed following the relevant guidelines and regulations of the Committee.

\section*{Funding}
All the authors acknowledge the support from EU HORIZON  2020 FET  Open  RIA  project (IBSEN) No. 662725. Kimmo Kaski acknowledges support from Academy of Finland Research project (COSDYN) No. 276439 and from the European Community's H2020 Program under the scheme ``INFRAIA-1-2014-2015: Research Infrastructures'', Grant agreement No. 654024 ``SoBigData: Social Mining and Big Data Ecosystem'' (http://www.sobigdata.eu). Daniel Monsivais acknowledges support from CONACYT, Mexico grant 383907. 

\section*{Competing interests}
The authors declare no competing interests.

\bibliographystyle{vancouver}

\begin{thebibliography}{10}

\bibitem{jennings2014human}
Jennings NR, Moreau L, Nicholson D, Ramchurn S, Roberts S, Rodden T, et~al.
\newblock Human-agent collectives.
\newblock Communications of the ACM. 2014;57(12):80--88.

\bibitem{van2017domo}
Van~Doorn J, Mende M, Noble SM, Hulland J, Ostrom AL, Grewal D, et~al.
\newblock Domo arigato Mr. Roboto: Emergence of automated social presence in
  organizational frontlines and customers’ service experiences.
\newblock Journal of Service Research. 2017;20(1):43--58.

\bibitem{bonabeau2002agent}
Bonabeau E.
\newblock Agent-based modeling: Methods and techniques for simulating human
  systems.
\newblock Proceedings of the National Academy of Sciences. 2002;99(suppl
  3):7280--7287.

\bibitem{kahneman2003maps}
Kahneman D.
\newblock Maps of bounded rationality: Psychology for behavioral economics.
\newblock American economic review. 2003;93(5):1449--1475.

\bibitem{groom2007can}
Groom V, Nass C.
\newblock Can robots be teammates?: Benchmarks in human--robot teams.
\newblock Interaction Studies. 2007;8(3):483--500.

\bibitem{shirado2017locally}
Shirado H, Christakis NA.
\newblock Locally noisy autonomous agents improve global human coordination in
  network experiments.
\newblock Nature. 2017;545(7654):370.

\bibitem{crandall2018cooperating}
Crandall JW, Oudah M, Ishowo-Oloko F, Abdallah S, Bonnefon JF, Cebrian M,
  et~al.
\newblock Cooperating with machines.
\newblock Nature communications. 2018;9(1):233.

\bibitem{kearns2012behavioral}
Kearns M, Judd S, Vorobeychik Y.
\newblock Behavioral experiments on a network formation game.
\newblock In: Proceedings of the 13th ACM Conference on Electronic Commerce.
  ACM; 2012. p. 690--704.

\bibitem{kearns2006experimental}
Kearns M, Suri S, Montfort N.
\newblock An experimental study of the coloring problem on human subject
  networks.
\newblock Science. 2006;313(5788):824--827.

\bibitem{judd2010behavioral}
Judd S, Kearns M, Vorobeychik Y.
\newblock Behavioral dynamics and influence in networked coloring and
  consensus.
\newblock Proceedings of the National Academy of Sciences.
  2010;107(34):14978--14982.

\bibitem{gale1962college}
Gale D, Shapley LS.
\newblock College admissions and the stability of marriage.
\newblock The American Mathematical Monthly. 1962;69(1):9--15.

\bibitem{laureti2003matching}
Laureti P, Zhang YC.
\newblock Matching games with partial information.
\newblock Physica A: Statistical Mechanics and its Applications.
  2003;324(1-2):49--65.

\bibitem{coviello2012human}
Coviello L, Franceschetti M, McCubbins MD, Paturi R, Vattani A.
\newblock Human matching behavior in social networks: an algorithmic
  perspective.
\newblock PloS one. 2012;7(8):e41900.

\bibitem{baronchelli2010modeling}
Baronchelli A, Gong T, Puglisi A, Loreto V.
\newblock Modeling the emergence of universality in color naming patterns.
\newblock Proceedings of the National Academy of Sciences.
  2010;107(6):2403--2407.

\bibitem{guazzini2015modeling}
Guazzini A, Vilone D, Donati C, Nardi A, Levnaji{\'c} Z.
\newblock Modeling crowdsourcing as collective problem solving.
\newblock Scientific reports. 2015;5:16557.

\bibitem{centola2015spontaneous}
Centola D, Baronchelli A.
\newblock The spontaneous emergence of conventions: An experimental study of
  cultural evolution.
\newblock Proceedings of the National Academy of Sciences.
  2015;112(7):1989--1994.

\bibitem{girvan2002community}
Girvan M, Newman ME.
\newblock Community structure in social and biological networks.
\newblock Proceedings of the national academy of sciences.
  2002;99(12):7821--7826.

\bibitem{wuchty2007increasing}
Wuchty S, Jones BF, Uzzi B.
\newblock The increasing dominance of teams in production of knowledge.
\newblock Science. 2007;316(5827):1036--1039.

\bibitem{lambiotte2009communities}
Lambiotte R, Panzarasa P.
\newblock Communities, knowledge creation, and information diffusion.
\newblock Journal of Informetrics. 2009;3(3):180--190.

\bibitem{lakhani2010topcoder}
Lakhani KR, Garvin DA, Lonstein E.
\newblock Topcoder (A): Developing software through crowdsourcing.
\newblock Harvard Business School General Management Unit Case No 610-032.
  2010;.

\bibitem{chauhan2018Schelling}
Chauhan A, Lenzner P, Molitor L.
\newblock Schelling Segregation with Strategic Agents.
\newblock In: Deng X, editor. Algorithmic Game Theory. Cham: Springer
  International Publishing; 2018. p. 137--149.

\bibitem{elkind2019schelling}
Elkind E, Gan J, Igarashi A, Suksompong W, Voudouris AA.
\newblock Schelling Games on Graphs.
\newblock arXiv preprint arXiv:190207937. 2019;.

\bibitem{schelling1971dynamic}
Schelling TC.
\newblock Dynamic models of segregation.
\newblock Journal of mathematical sociology. 1971;1(2):143--186.

\bibitem{dreze1980hedonic}
Dreze JH, Greenberg J.
\newblock Hedonic coalitions: Optimality and stability.
\newblock Econometrica (pre-1986). 1980;48(4):987.

\bibitem{aziz2016hedonic}
Aziz H, Savani R.
\newblock Hedonic Games.
\newblock In: Brandt F, Procaccia AD, editors. Handbook of computational social
  choice. New York: Cambridge University Press; 2016. p. 356--377.

\bibitem{otree}
Chen DL, Schonger M, Wickens C.
\newblock OTree - An Open-Source Platform for Laboratory, Online and Field
  Experiments.
\newblock Journal of Behavioral and Experimental Finance. 2016;9:88--97.

\bibitem{kleinberg2000small}
Kleinberg J.
\newblock The small-world phenomenon: An algorithmic perspective.
\newblock In: Proceedings of the thirty-second annual ACM symposium on Theory
  of computing. ACM; 2000. p. 163--170.

\bibitem{halpern1990knowledge}
Halpern JY, Moses Y.
\newblock Knowledge and common knowledge in a distributed environment.
\newblock Journal of the ACM (JACM). 1990;37(3):549--587.

\bibitem{vulkan2000economist}
Vulkan N.
\newblock An economist's perspective on probability matching.
\newblock Journal of economic surveys. 2000;14(1):101--118.

\bibitem{shanks2002re}
Shanks DR, Tunney RJ, McCarthy JD.
\newblock A re-examination of probability matching and rational choice.
\newblock Journal of Behavioral Decision Making. 2002;15(3):233--250.

\bibitem{arganda2012common}
Arganda S, P{\'e}rez-Escudero A, de~Polavieja GG.
\newblock A common rule for decision making in animal collectives across
  species.
\newblock Proceedings of the National Academy of Sciences.
  2012;109(50):20508--20513.

\bibitem{vul2014one}
Vul E, Goodman N, Griffiths TL, Tenenbaum JB.
\newblock One and done? Optimal decisions from very few samples.
\newblock Cognitive science. 2014;38(4):599--637.

\bibitem{krafft2016modeling}
Krafft PM, Baker CL, Pentland AS, Tenenbaum JB.
\newblock Modeling human ad hoc coordination.
\newblock In: Proceedings of the Thirtieth AAAI Conference on Artificial
  Intelligence. AAAI Press; 2016. p. 3740--3746.

\bibitem{blume2001interactions}
Blume L, Durlauf SN.
\newblock The interactions-based approach to socioeconomic behavior.
\newblock Social dynamics. 2001;15.

\bibitem{young2015evolution}
Young HP.
\newblock The evolution of social norms.
\newblock economics. 2015;7(1):359--387.

\bibitem{doebeli2005models}
Doebeli M, Hauert C.
\newblock Models of cooperation based on the Prisoner's Dilemma and the
  Snowdrift game.
\newblock Ecology letters. 2005;8(7):748--766.

\bibitem{nowak2006five}
Nowak MA.
\newblock Five rules for the evolution of cooperation.
\newblock science. 2006;314(5805):1560--1563.

\bibitem{galeotti2010network}
Galeotti A, Goyal S, Jackson MO, Vega-Redondo F, Yariv L.
\newblock Network games.
\newblock The review of economic studies. 2010;77(1):218--244.

\bibitem{helbing2009outbreak}
Helbing D, Yu W.
\newblock The outbreak of cooperation among success-driven individuals under
  noisy conditions.
\newblock Proceedings of the National Academy of Sciences.
  2009;106(10):3680--3685.

\bibitem{suri2011cooperation}
Suri S, Watts DJ.
\newblock Cooperation and contagion in web-based, networked public goods
  experiments.
\newblock PloS one. 2011;6(3):e16836.

\bibitem{wang2012cooperation}
Wang J, Suri S, Watts DJ.
\newblock Cooperation and assortativity with dynamic partner updating.
\newblock Proceedings of the National Academy of Sciences.
  2012;109(36):14363--14368.

\bibitem{antonioni2015short}
Antonioni A, Tomassini M, S{\'a}nchez A.
\newblock Short-range mobility and the evolution of cooperation: an
  experimental study.
\newblock Scientific reports. 2015;5.

\bibitem{thomas2014psychology}
Thomas KA, DeScioli P, Haque OS, Pinker S.
\newblock The psychology of coordination and common knowledge.
\newblock Journal of personality and social psychology. 2014;107(4):657.

\bibitem{lancichinetti2008benchmark}
Lancichinetti A, Fortunato S, Radicchi F.
\newblock Benchmark graphs for testing community detection algorithms.
\newblock Physical review E. 2008;78(4):046110.

\bibitem{granovetter1977strength}
Granovetter MS.
\newblock The strength of weak ties.
\newblock In: Social networks. Elsevier; 1977. p. 347--367.

\bibitem{larsen2004bonding}
Larsen L, Harlan SL, Bolin B, Hackett EJ, Hope D, Kirby A, et~al.
\newblock Bonding and bridging: Understanding the relationship between social
  capital and civic action.
\newblock Journal of Planning Education and Research. 2004;24(1):64--77.

\bibitem{ellison2007benefits}
Ellison NB, Steinfield C, Lampe C.
\newblock The benefits of Facebook ``friends:'' Social capital and college
  students' use of online social network sites.
\newblock Journal of computer-mediated communication. 2007;12(4):1143--1168.

\bibitem{brummitt2015jigsaw}
Brummitt CD, Chatterjee S, Dey PS, Sivakoff D, et~al.
\newblock Jigsaw percolation: What social networks can collaboratively solve a
  puzzle?
\newblock The Annals of Applied Probability. 2015;25(4):2013--2038.

\bibitem{saad2009coalitional}
Saad W, Han Z, Debbah M, Hjorungnes A, Basar T.
\newblock Coalitional game theory for communication networks.
\newblock IEEE Signal Processing Magazine. 2009;26(5):77--97.

\end{thebibliography}

\begin{thebibliography}{99}
\bibitem{zar} Zar JH (2013) Biostatistical analysis: Pearson new international edition. Pearson Education, Upper Saddle River.
\bibitem{agresti} Agresti, A., \& Kateri, M. (2011). Categorical data analysis (pp. 489-533). Springer Berlin Heidelberg.
\bibitem{lme4} Bates, D. \emph{et al.} (2012). Package `lme4'. CRAN. R Foundation for Statistical Computing, Vienna, Austria.
\end{thebibliography}

\clearpage
\setcounter{figure}{0}
\setcounter{section}{0}
\renewcommand{\thesection}{S\arabic{section}}  
\renewcommand{\thetable}{S\arabic{table}}  
\renewcommand{\thefigure}{S\arabic{figure}} 

{\it Supplementary Material}\\
\section{Model details}
\begin{figure*}[h]
\centering
\includegraphics[width=6.6cm]{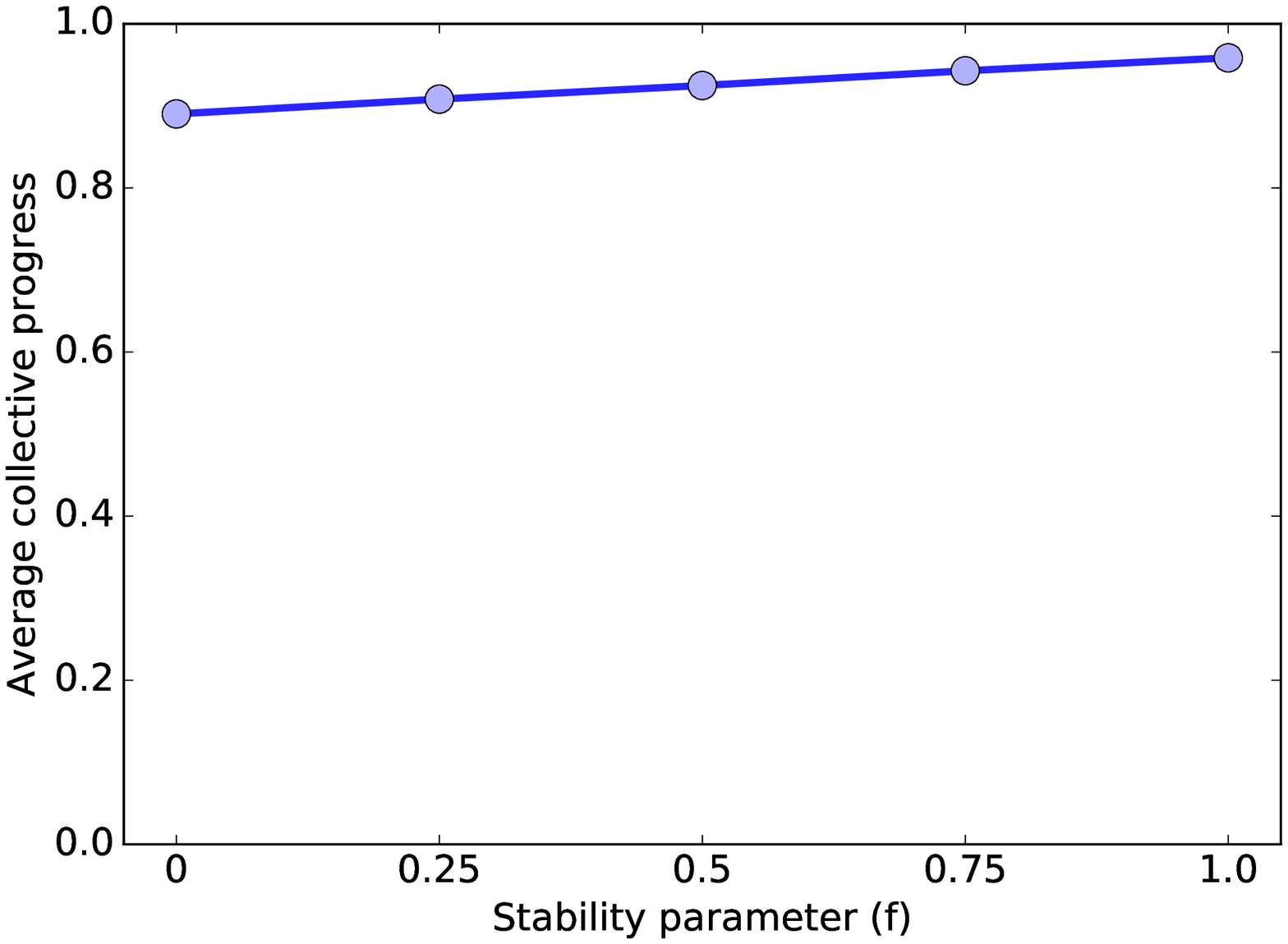}
\includegraphics[width=6.6cm]{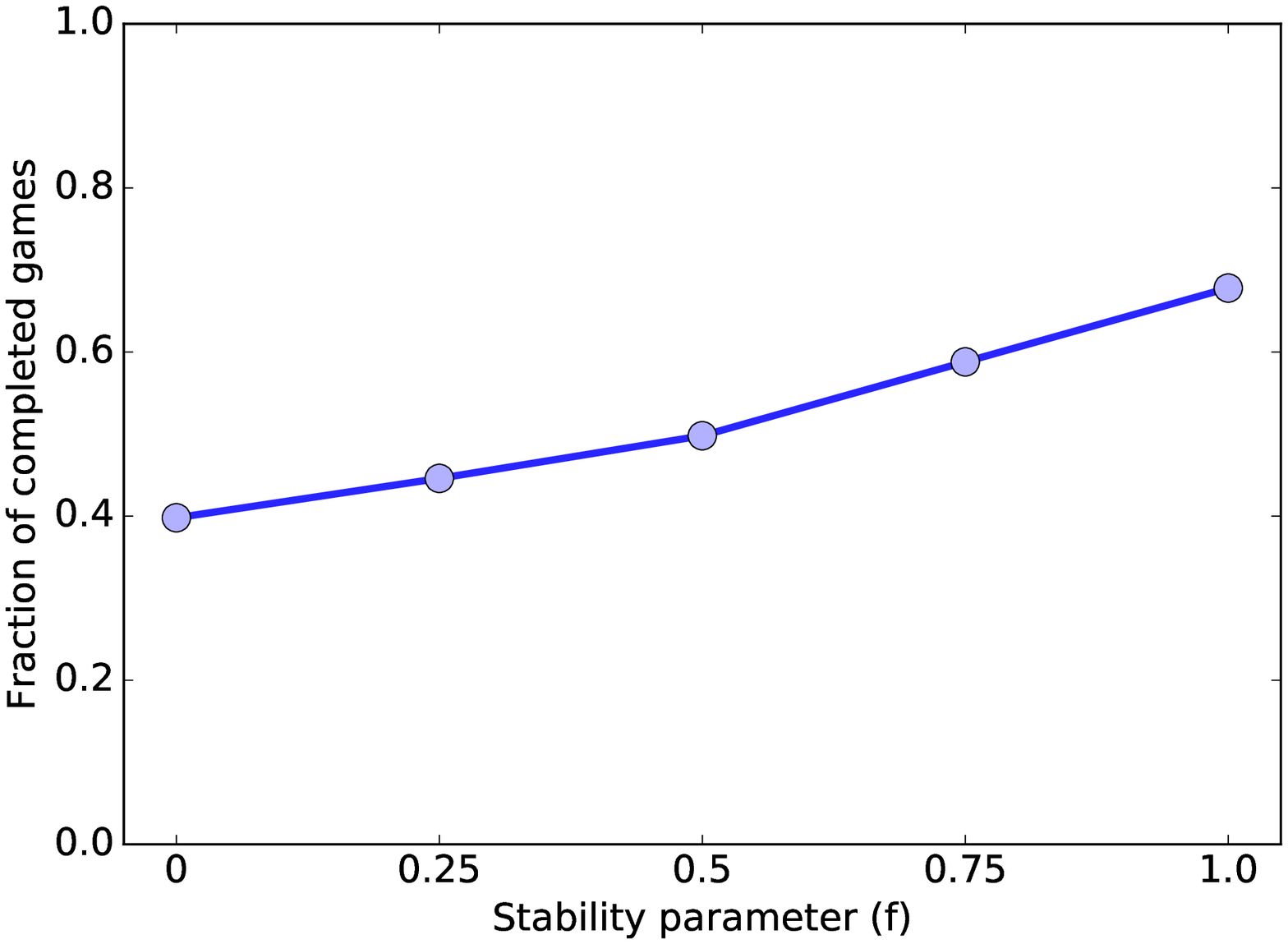}
\caption{The effect of the stability rule on the average collective progress (left) and the fraction of games completed (right). The parameter $f$ allows us to scan between the limits of having no stability rule ($f=0$) and having absolute implementation of the rule ($f=1$).}
\label{figs1}
\end{figure*}

\begin{figure*}[h]
\centering
\includegraphics[width=6.6cm]{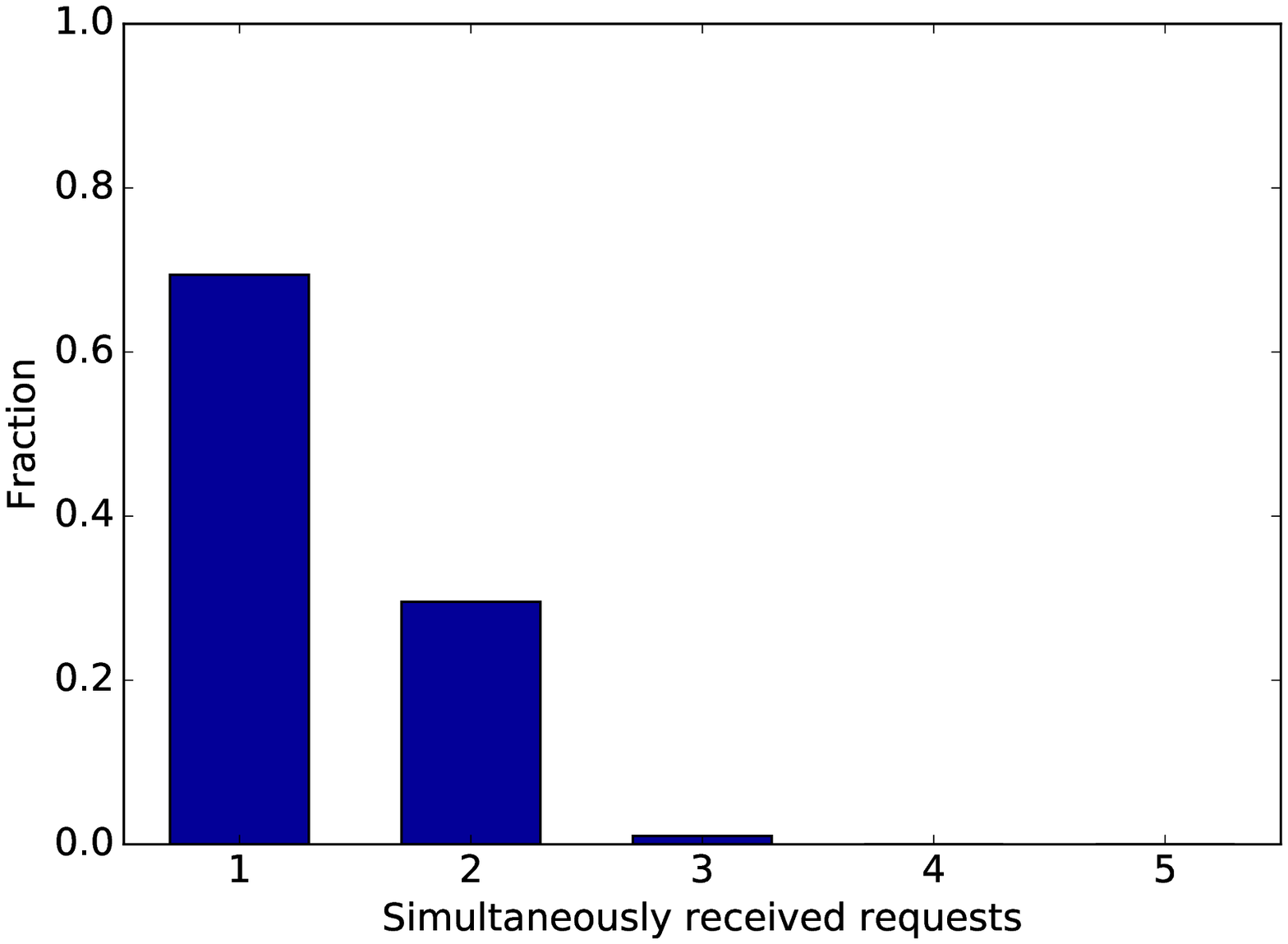}
\includegraphics[width=6.6cm]{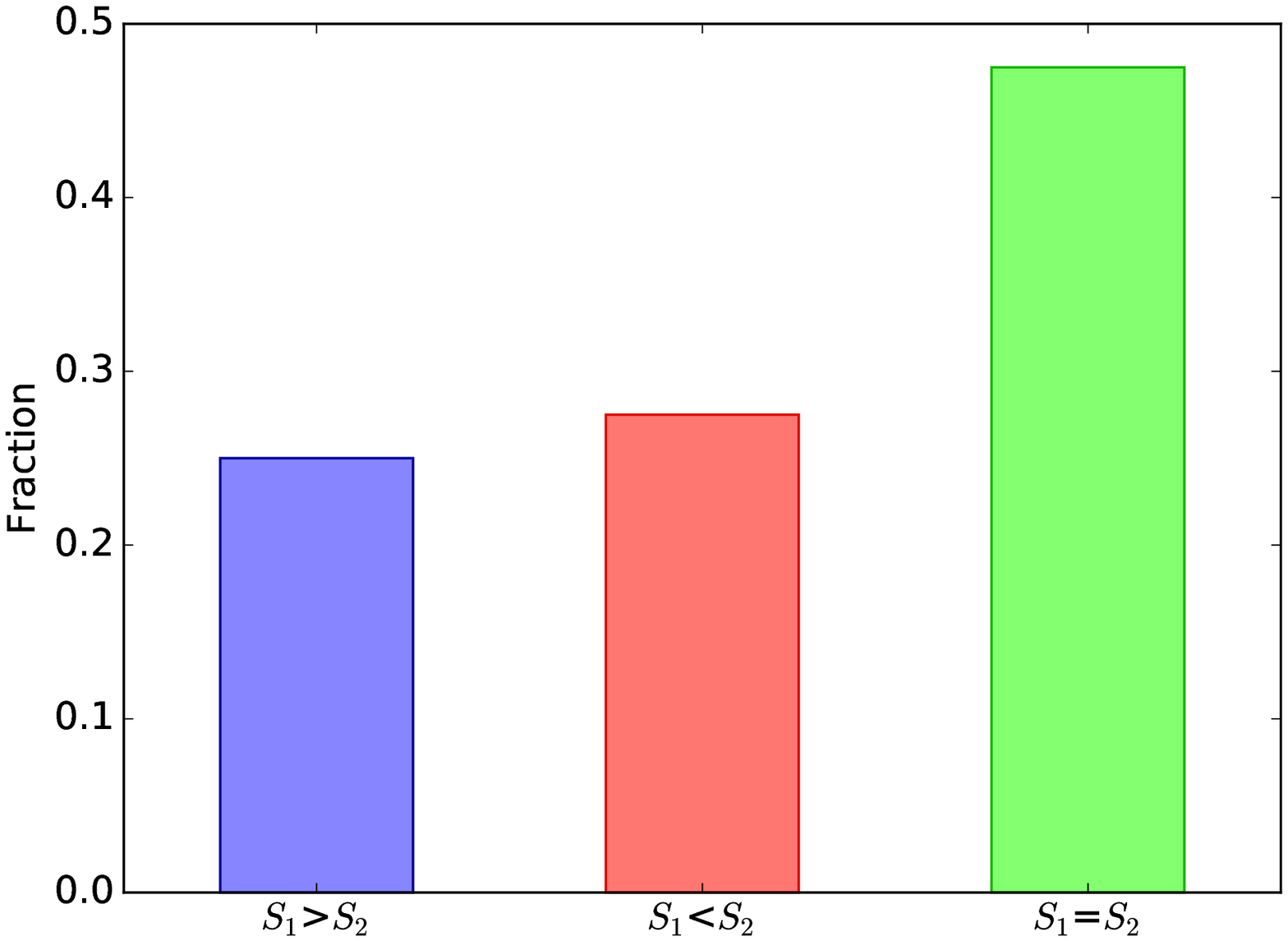}
\caption{(Left) The distribution of the number of simultaneous requests received by a player. (Right) The distribution of accepted requests when receiving two simultaneous requests, where $S_1$ denotes the cluster size of the selected requester and $S_2$ denotes the cluster size of the denied requester.}
\label{figs2}
\end{figure*}

The Fig.~\ref{figs1} illustrates the stability rule implemented in the Model (main text). The Fig.~\ref{figs2} shows the statistics of simultaneous requests and the nature of acceptances.

\section{Testing for learning during games}
\begin{figure*}[h]
\includegraphics[width=15cm]{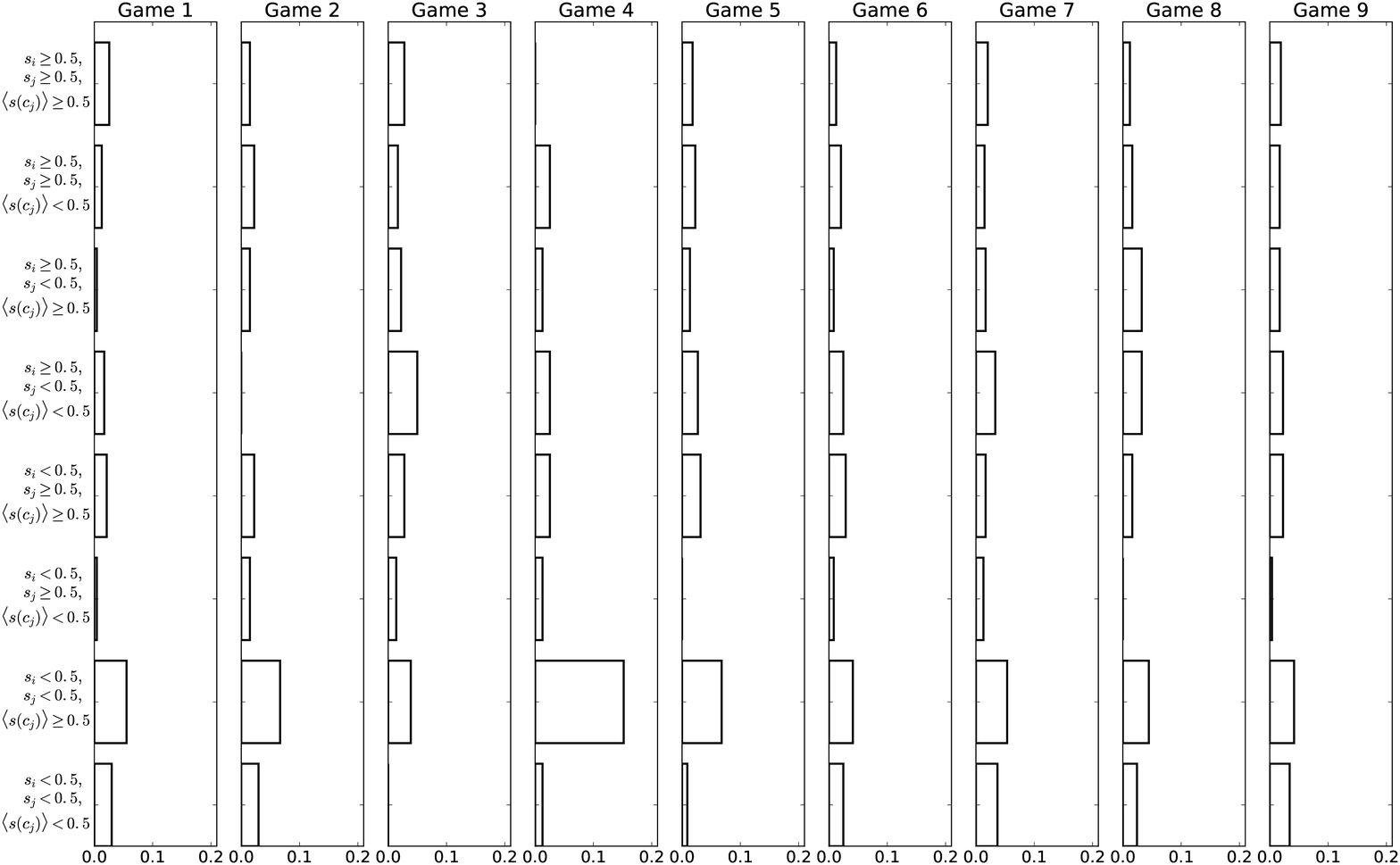}
\caption{Requesting decisions of the subjects on each one of the 9 games. The decisions are categorized into 8 different strategies and the fraction of cases in each strategy is shown for each game. The counts for the decisions in a given game is normalized using the total decisions in the game. }
\label{fig-S1}
\end{figure*}

\begin{figure*}[h]
\includegraphics[width=15cm]{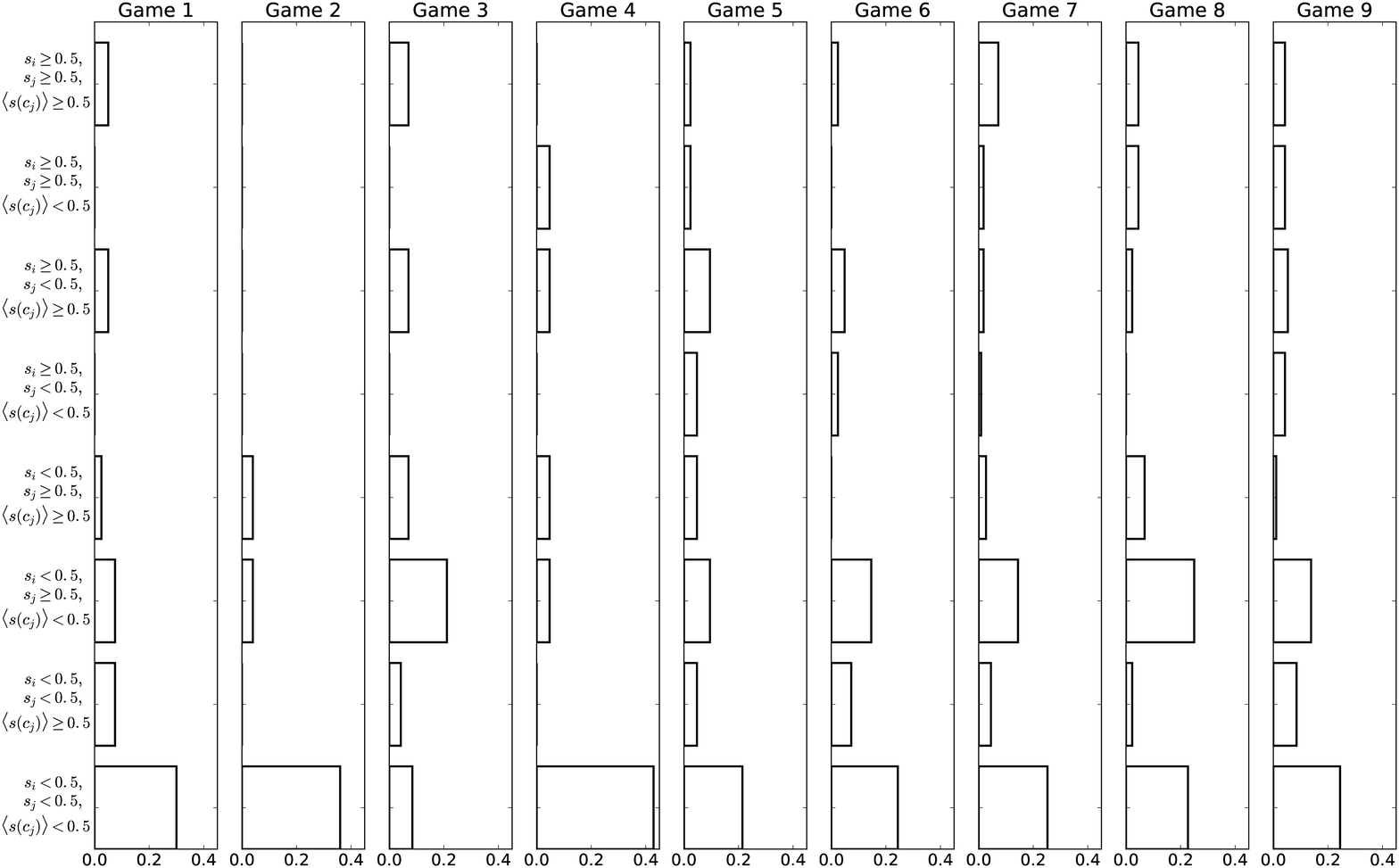}
\caption{ Accepting decisions of the subjects during the 9 games. The decisions are categorized into 8 different strategies and the fraction of cases in each strategy is shown for the different games.  
}
\label{fig-S2}
\end{figure*}

The model is able to describe (provide probabilities on) the choices made by the players in terms of the three variables of the system -- (i) cluster size of the focal player, $s_i$ , (ii) cluster size of the neighbour to whom the request is made or whose request is received, $s_j$ , and (iii) the average of the cluster sizes of the number of neighbours $n_i(c_j)$  of player $i$ that have the same colour as $j$, $\langle s(c_j)\rangle$. To investigate whether the subjects were learning and changing their preferences over the span of the 9 games, we utilize the aforementioned information and first categorize the observed decisions in terms of the three variables $s_i$, $s_j$, and $\langle s(c_j)\rangle$. We divide the 3-dimensional decision space into octants ($s_i \gtreqless 0.5, s_j\gtreqless 0.5, \langle s(c_j)\rangle \gtreqless 0.5$) and count the number of choices in each of these 8 octants (categories) in a given game. We then compare the distribution of the choices across the 9 different games. The distribution of choices for requesters and acceptors are shown in Fig.~\ref{fig-S1} and Fig.~\ref{fig-S2}, respectively. We compare the distribution of counts using the Pearson's chi-squared test. We also clustered the games into three (games 1 to 3, 4 to 6, and 7 to 9) and two (games 1 to 5,  and 6 to 9) sets by aggregating the counts of the individual games. In Table~\ref{tab-s1} we show the results the test with the strategy being the categorical variable and the null hypothesis being that there is no difference between the distribution of strategy across the groups (games) \cite{zar}. We find that $p$-values are not small enough to reject the null hypothesis. Therefore, we conclude that no significant learning behaviour was observed.

\begin{table*}[]
\caption{ Distribution of strategies in different sets of games are tested for difference using the Pearson's chi-squared test.}

\begin{tabular}{|c|c|c|c|c|c|}
\hline
{Groups}                                                                   &    & \multicolumn{2}{c|}{Requester}        & \multicolumn{2}{c|}{Acceptor}         \\ \cline{2-6} 
                                                                                         & df & $\chi^2$ & p-value & $\chi^2$ & p-value \\ \hline
All 9 games                                                                              & 56 & 67.01                       & 0.15    & 67.63               & 0.14    \\ \hline
\begin{tabular}[c]{@{}c@{}}Grp-1=g1+g2+g3\\ Grp-2=g5+g7+g6\\ Grp-3=g7+g8+g9\end{tabular} & 14 & 20.37                       & 0.12    & 16.26                       & 0.30    \\ \hline
\begin{tabular}[c]{@{}c@{}}Grp-1=g1+g2+g3+g4+g5\\ Grp-2=g6+g7+g8+g9\end{tabular}      & 7  & 13.17                       & 0.07    & 8.54                        & 0.29    \\ \hline
\end{tabular}
\label{tab-s1}
\end{table*}

\section{Testing for behavioural heterogeneity}

Below we provide a method to study and model the aspect of heterogeneity among agents. In the basic modelling scheme (refer to the main text, Model section) we cast the model in the form of a logistic regression that learn the parameters $\alpha$, $\beta$ and $\delta$ from the data. We extend the scheme to consider mixed effects logistic regression \cite{agresti}, where an additional variable for subjects account for random effects (heterogeneity). We consider the following four models with different levels of complexity that we implement using an R programming package for generalized linear mixed models \cite{lme4}:

\begin{align*}
\textrm{Model-0: }\textrm{logit}(P_\omega^{(m,k)} / P_0^{(m,k)})&=\lambda+\alpha s_i^{(m,k)}+\beta s_j^{(m,k)}+\delta \langle s(c_j)\rangle^{(m,k)}\\\    
\textrm{Model-1: }\textrm{logit}(P_\omega^{(m,k)} / P_0^{(m,k)})&=(\lambda+\lambda_m)+\alpha s_i^{(m,k)}+\beta s_j^{(m,k)}+\delta \langle s(c_j)\rangle^{(m,k)}\\\
\textrm{Model-2: }\textrm{logit}(P_\omega^{(m,k)} / P_0^{(m,k)})&=\lambda+(\alpha+ \alpha_m)s_i^{(m,k)}+(\beta+\beta_m) s_j^{(m,k)}+(\delta+\delta_m) \langle s(c_j)\rangle^{(m,k)}\\\    
\textrm{Model-3: }\textrm{logit}(P_\omega^{(m,k)} / P_0^{(m,k)})&=(\lambda+\lambda_m)+(\alpha+ \alpha_m)s_i^{(m,k)}+(\beta+\beta_m) s_j^{(m,k)}+(\delta+\delta_m) \langle s(c_j)\rangle^{(m,k)}.\\\
\end{align*}
In the above equations the superscript $(m,k)$ refers to the $k$-th observation for the $m$-th subject in a game. Model-0 is the basic model used without random effects. The intercept $\lambda_m$ and the coefficients $\alpha_m$, $\beta_m$ and $\delta_m$ account for random effects due to the subjects. Model-1 is the random intercept model assuming that different subjects tend to have different intercepts and that slopes do not differ. Model-2 ignores random effects in the intercept and considers the slopes. Model-3 is a combination of Model-1 and Model-2. Using the data from the requesters' decisions we fit the models. In table~\ref{tab-s2} we provide the intercepts, coefficients and their standard deviations obtained from the models (all the coefficients are significant with, $p$-value $\leq$0.001).  An Anova test shows that out of the 4 models, Model-1 and Model-3 are the best candidates (AIC$\simeq$572 and $p<$0.001). We also considered models of higher complexity, for instance, with correlated intercepts and slopes, but on these the optimization method failed to converge. Using this scheme we are able to obtain the standard deviations of the intercepts and coefficients that quantify the heterogeneity of the subjects. In principle, the standard deviations can be used for generating randomly distributed intercepts and coefficients in the agent-based model. In the case of the acceptors the sparseness of the data (accepting instances are much lower than requesting instances) the fits could not be performed. However, from the similarity between requesting and accepting coefficients, as observed in Table-1 of the main text, we may consider the standard deviations in the accepting coefficients to be of similar magnitude.

\begin{table*}[]
\caption{Results from mixed effects logistic regression models from the data of the requesters. The intercepts, coefficients and their standard deviations obtained from the models are shown. Model-0 does not have any mixed effects. Feature scaling of the input to models have been performed to enhance the optimization.}

\begin{tabular}{|@{\hskip 0.1in}c@{\hskip 0.1in}|@{\hskip 0.1in}c@{\hskip 0.1in}|@{\hskip 0.1in}c@{\hskip 0.1in}|@{\hskip 0.1in}c@{\hskip 0.1in}|@{\hskip 0.1in}c@{\hskip 0.1in}|@{\hskip 0.1in}c@{\hskip 0.1in}|@{\hskip 0.1in}c@{\hskip 0.1in}|@{\hskip 0.1in}c@{\hskip 0.1in}|@{\hskip 0.1in}c@{\hskip 0.1in}|}
\hline
Model   & $\lambda$ & $\alpha$ & $\beta$  & $\delta$ & $\sigma_{\lambda}$ & $\sigma_{\alpha}$ & $\sigma_{\beta}$ & $\sigma_{\delta}$\\ \hline
Model-0 & -0.97  & -1.36 & -1.55 & 1.05  &              &             &            &             \\ \hline
Model-1 & -1.12  & -1.65 & -1.65 & 1.13  & 1.08         &             &            &             \\ \hline
Model-2 & -1.05  & -1.72 & -1.71 & 1.07  &              & 0.86        & 0.41       & 0.61        \\ \hline
Model-3 & -1.21  & -1.99 & -1.93 & 1.17  & 1.17         & 0.87        & 0.47       & 0.52        \\ \hline
\end{tabular}
\label{tab-s2}

\end{table*}
\clearpage
\section{Visual presentation given to participants before the experiment}
\begin{figure*}[h!]
\centering
\fbox{\includegraphics[page=1,width=0.45\textwidth] {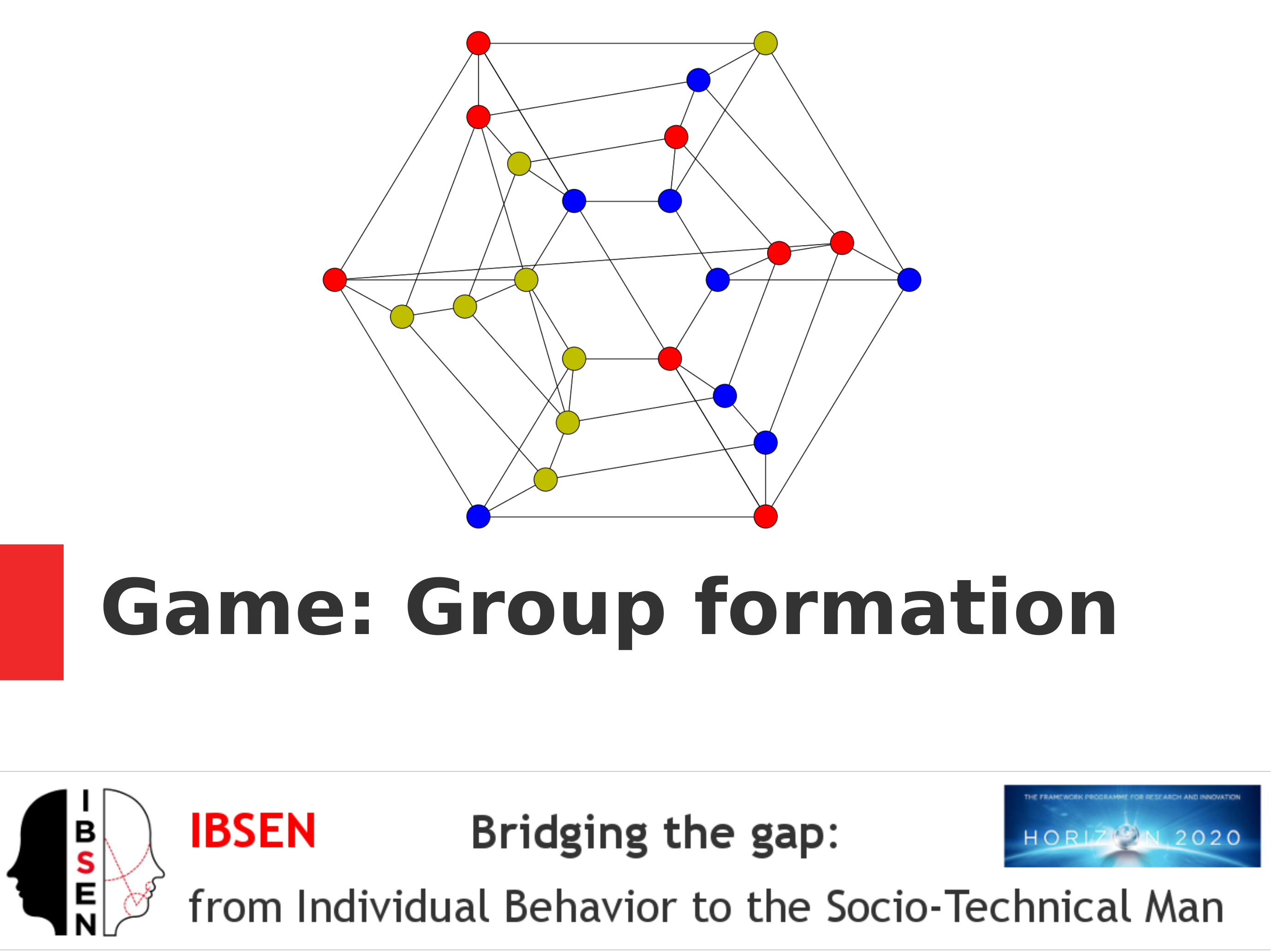}}
\fbox{\includegraphics[page=2,width=0.45\textwidth] {IntroPresis.pdf}}
\fbox{\includegraphics[page=3,width=0.45\textwidth] {IntroPresis.pdf}}
\fbox{\includegraphics[page=4,width=0.45\textwidth] {IntroPresis.pdf}}
\fbox{\includegraphics[page=5,width=0.45\textwidth] {IntroPresis.pdf}}
\fbox{\includegraphics[page=6,width=0.45\textwidth] {IntroPresis.pdf}}
\caption{Slides 1-6 of the visual material (presentation) given to the participants before the experiment to introduce them to rules, goal and dynamics of the game on which they participated. }
\label{fig:Pres1}
\end{figure*}
\clearpage
\newpage

\begin{figure*}[h!]
\centering
\fbox{\includegraphics[page=7,width=0.45\textwidth] {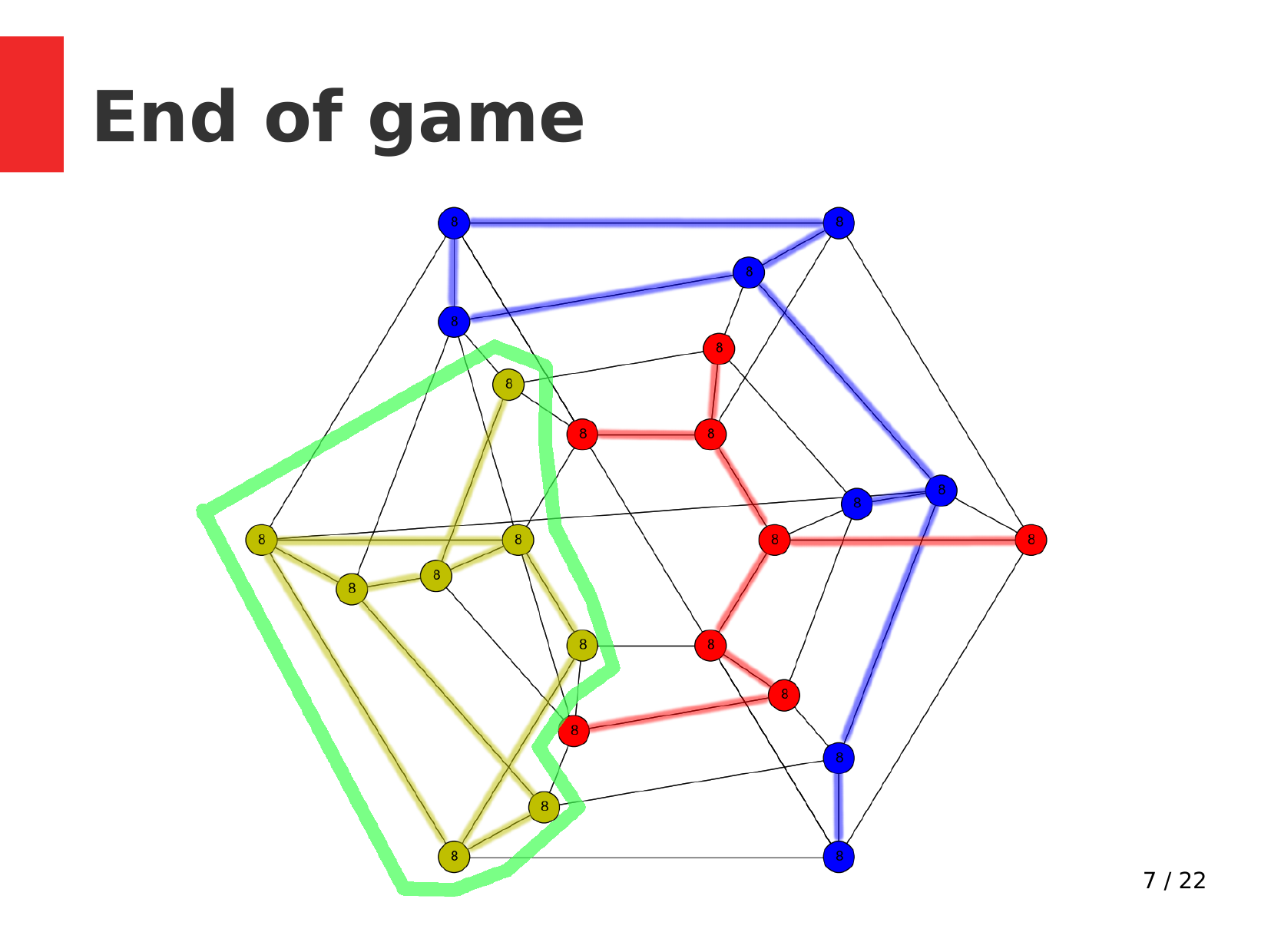}}
\fbox{\includegraphics[page=8,width=0.45\textwidth] {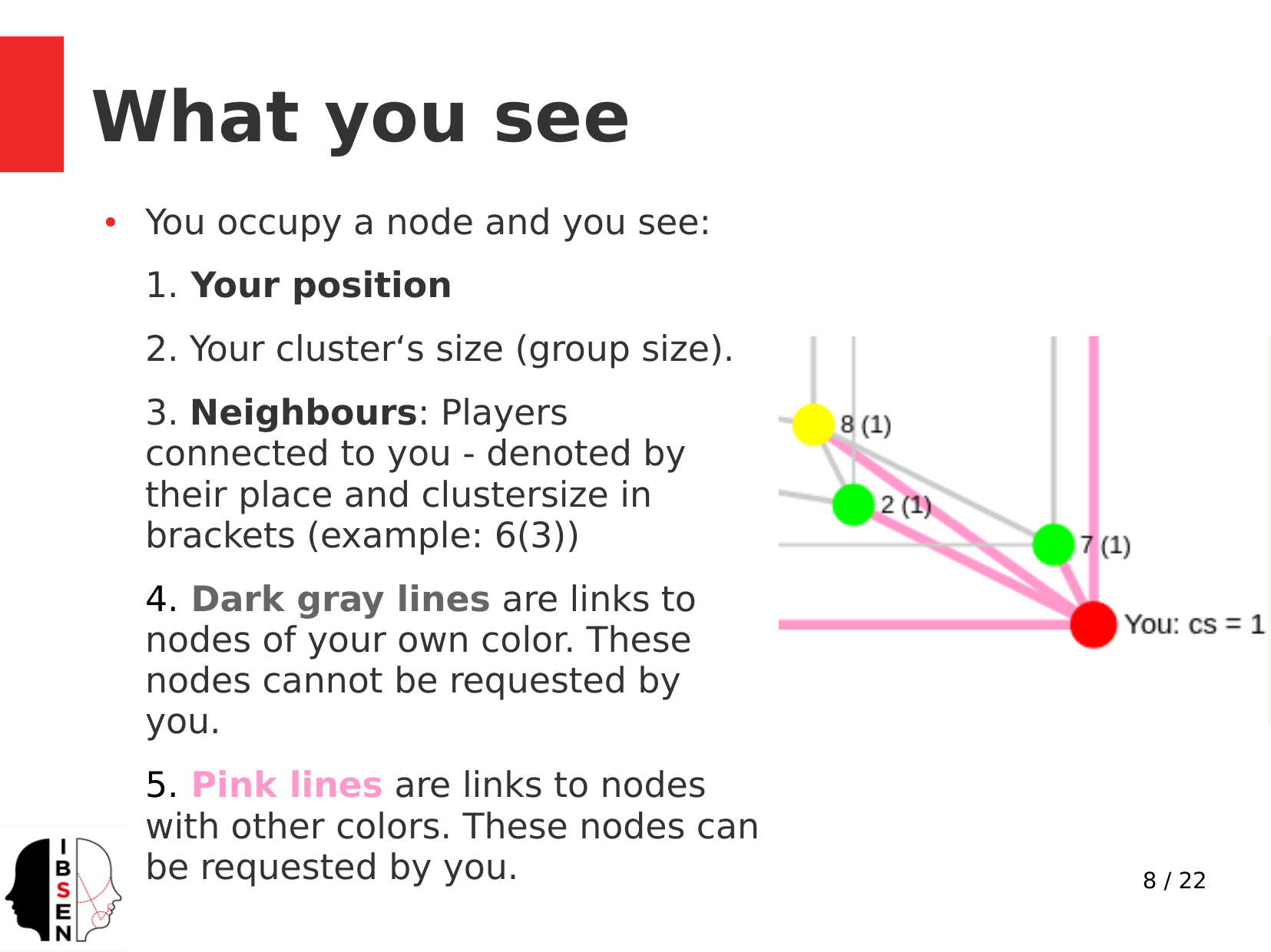}}
\fbox{\includegraphics[page=9,width=0.45\textwidth] {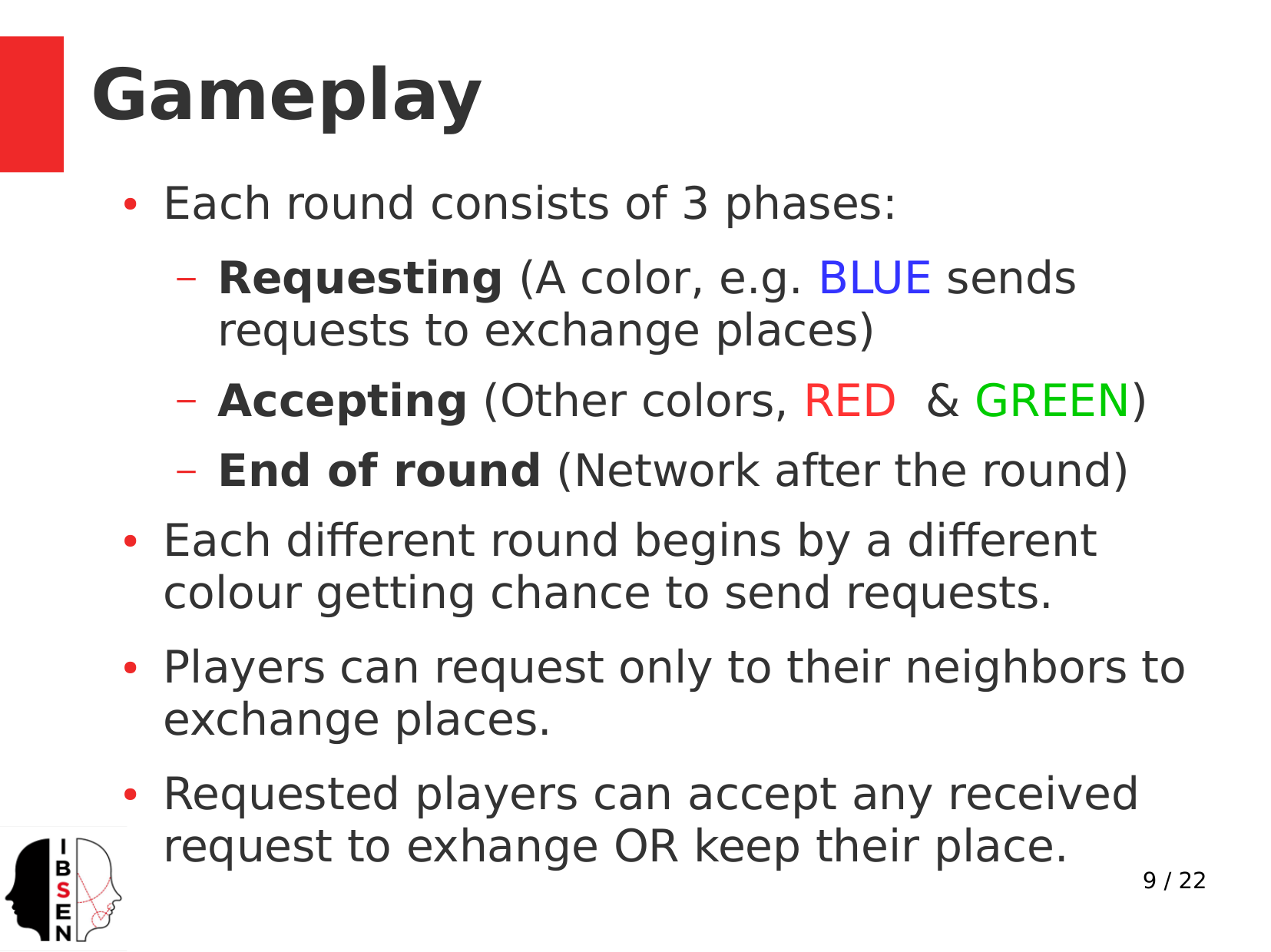}}
\fbox{\includegraphics[page=10,width=0.45\textwidth] {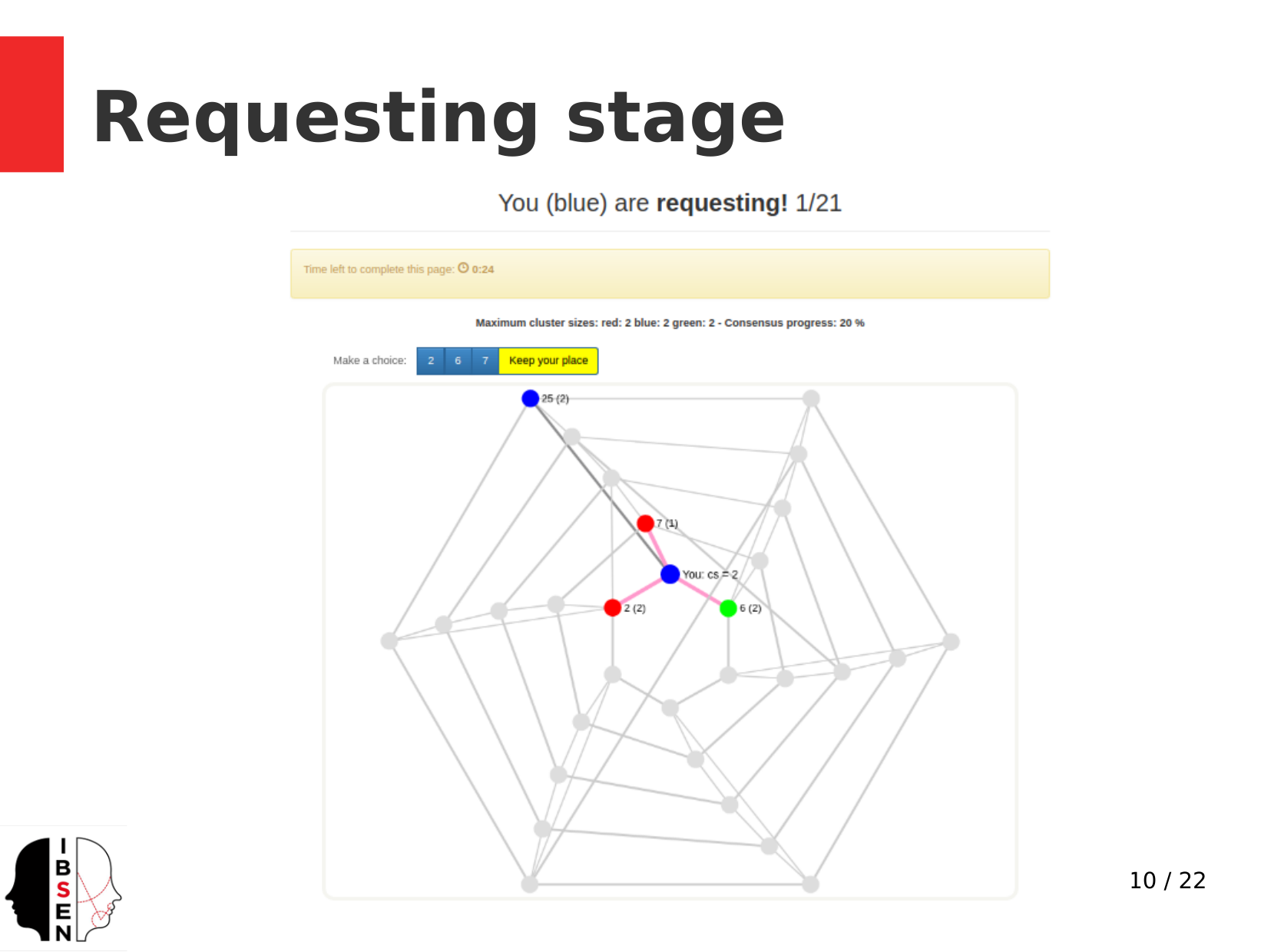}}
\fbox{\includegraphics[page=11,width=0.45\textwidth] {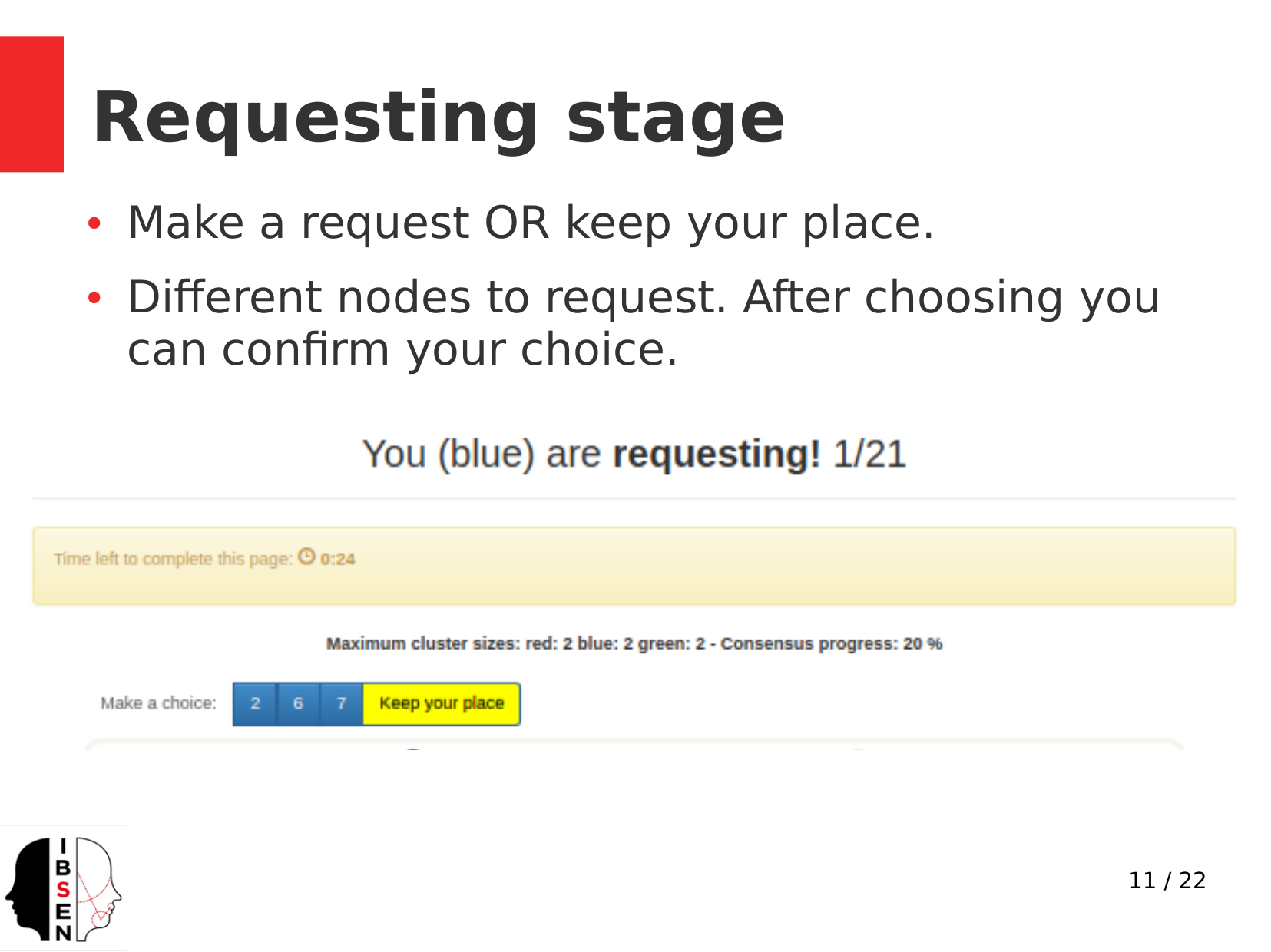}}
\fbox{\includegraphics[page=12,width=0.45\textwidth] {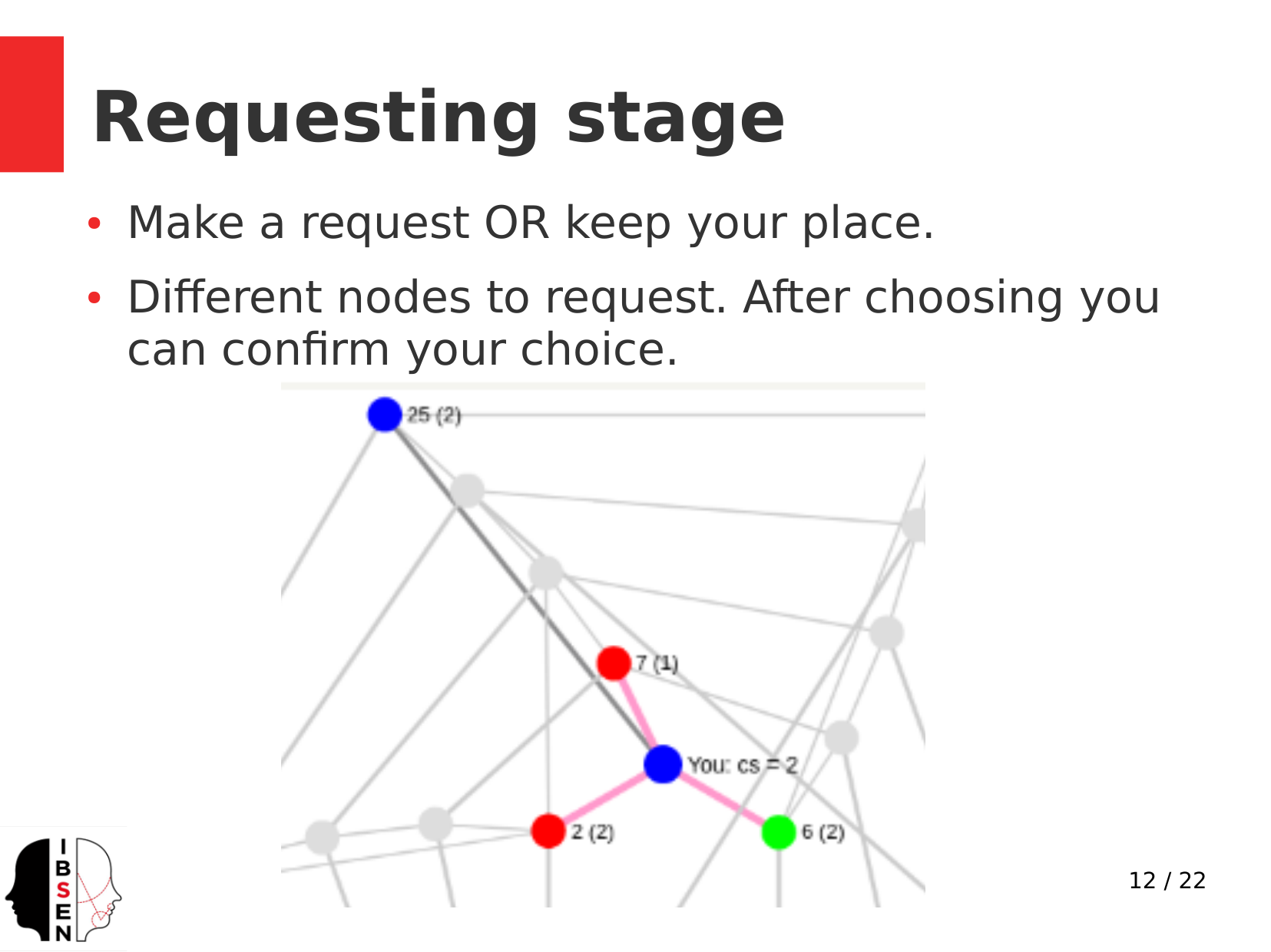}}
\caption{Slides 13-18 of the visual material (presentation) given to the participants. }
\label{fig:Pres2}
\end{figure*}
\clearpage
\newpage

\begin{figure*}[h!]
\centering
\fbox{\includegraphics[page=13,width=0.45\textwidth] {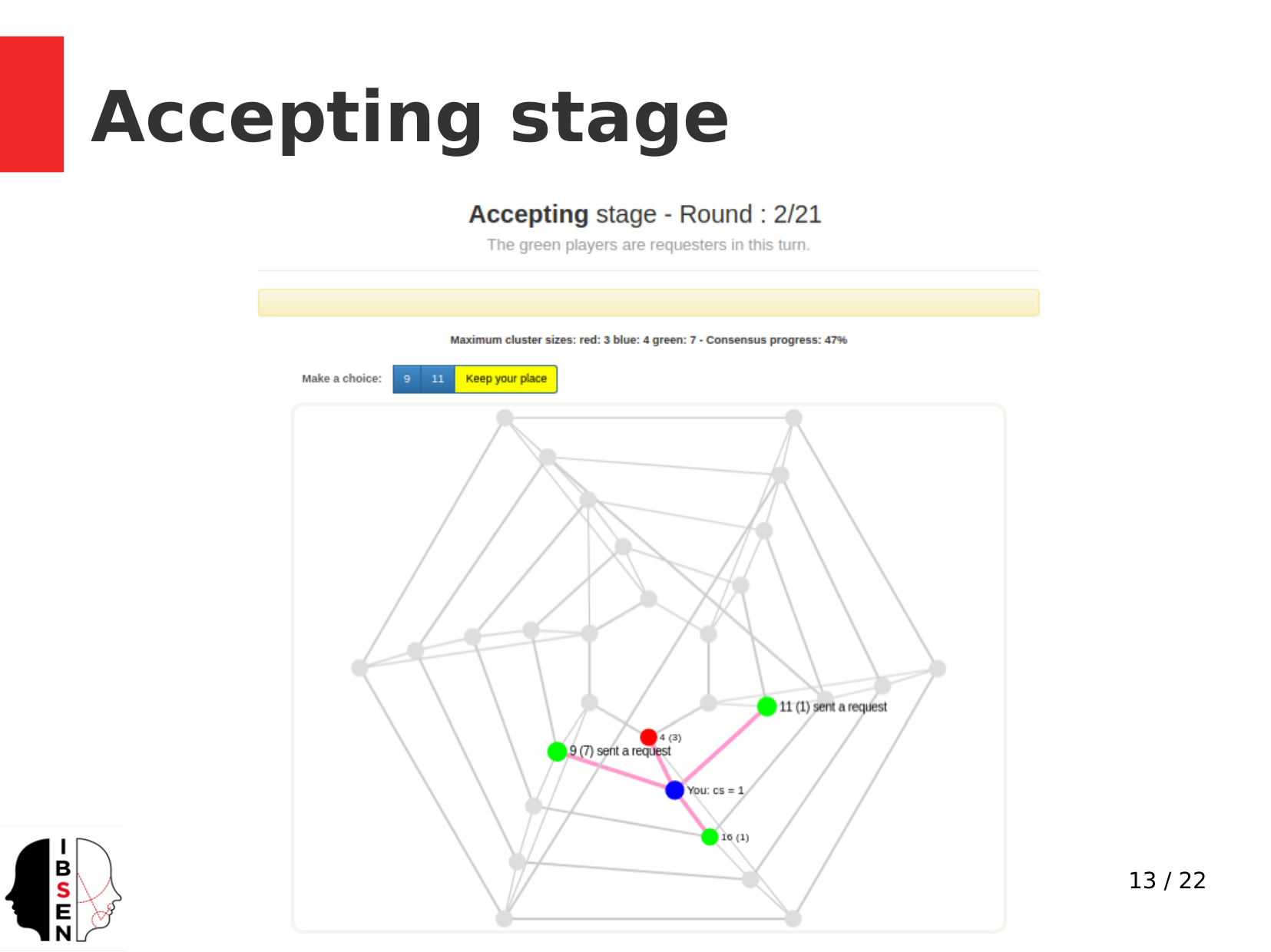}}
\fbox{\includegraphics[page=14,width=0.45\textwidth] {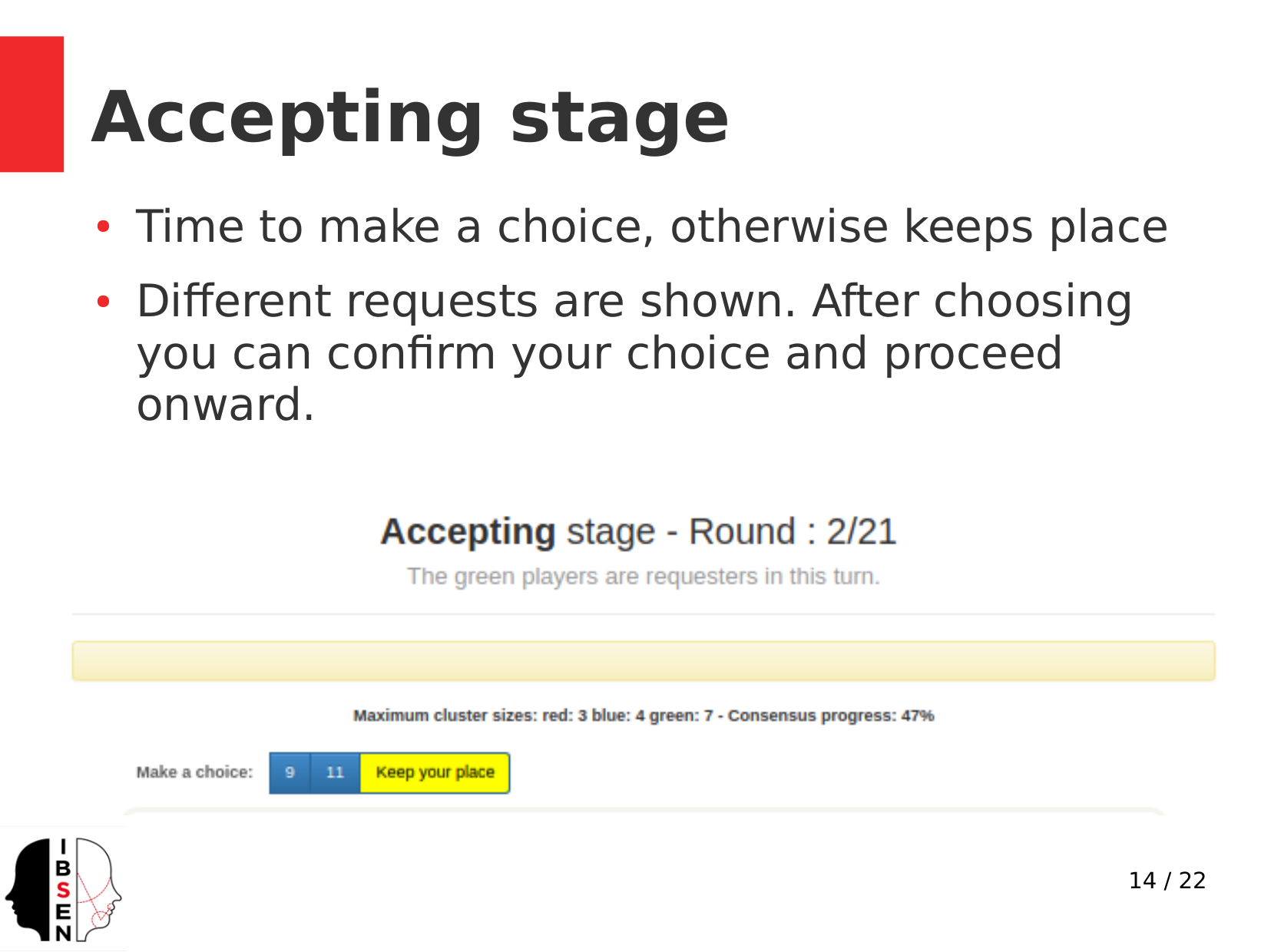}}
\fbox{\includegraphics[page=15,width=0.45\textwidth] {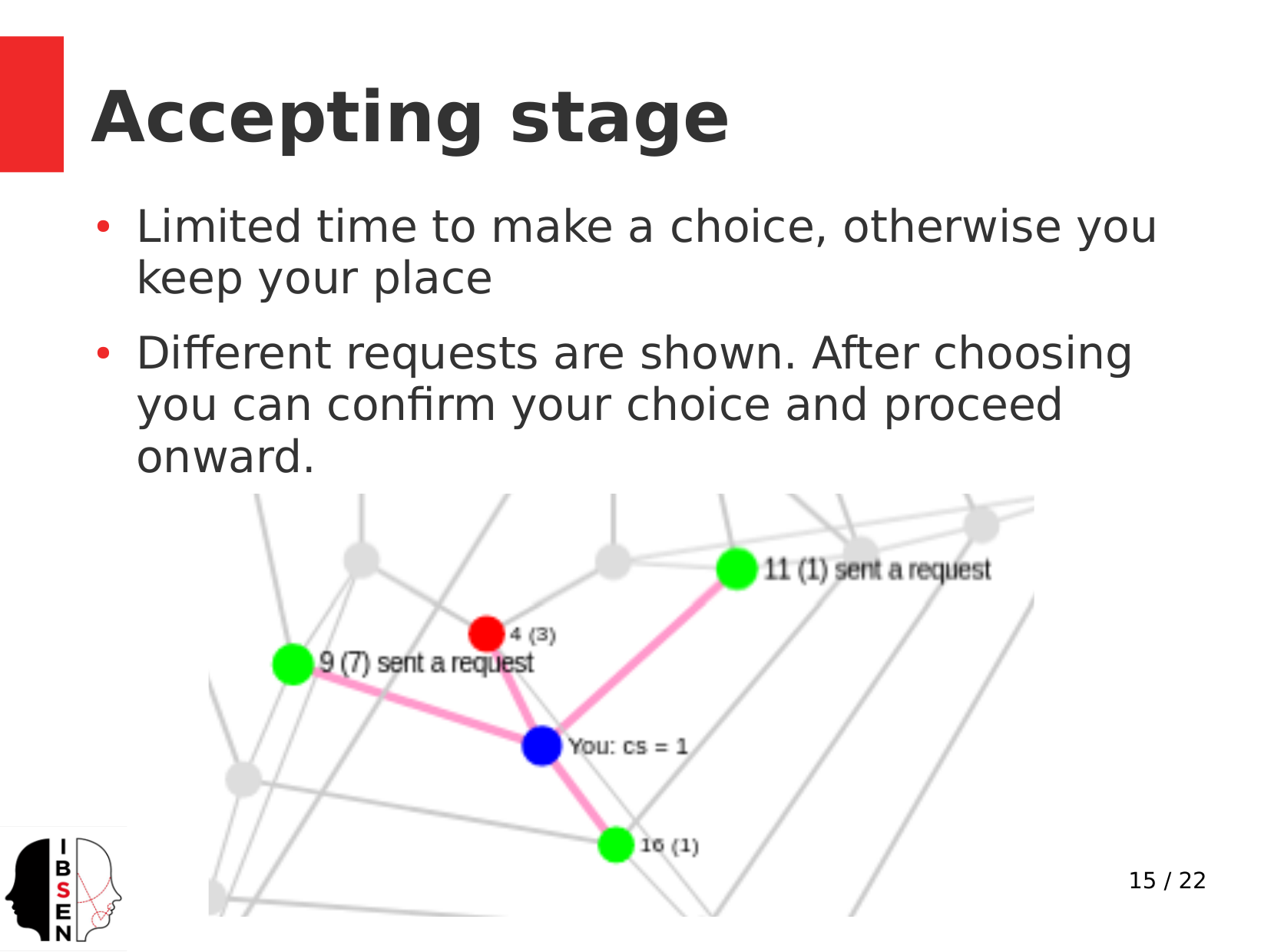}}
\fbox{\includegraphics[page=16,width=0.45\textwidth] {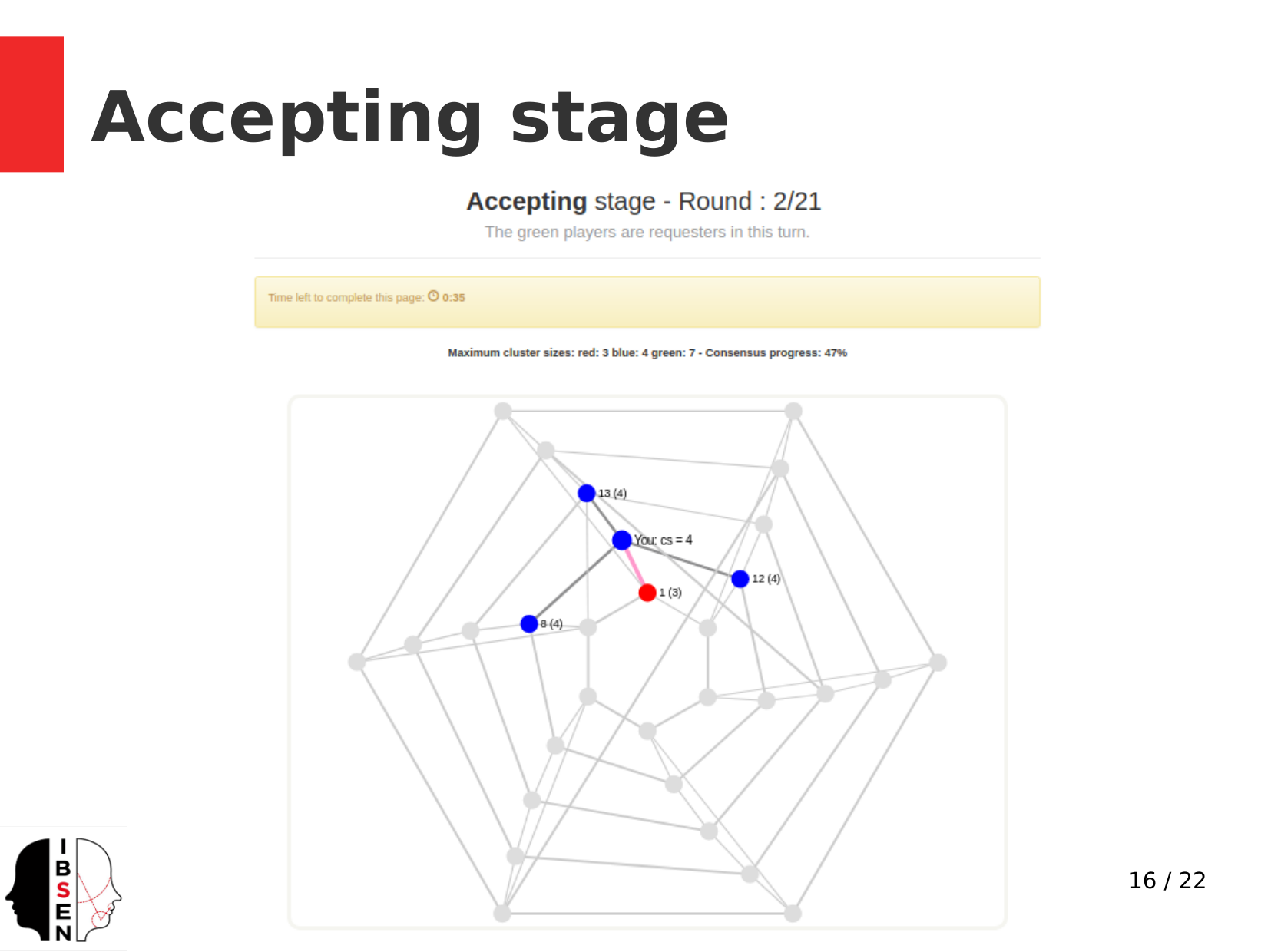}}
\fbox{\includegraphics[page=17,width=0.45\textwidth] {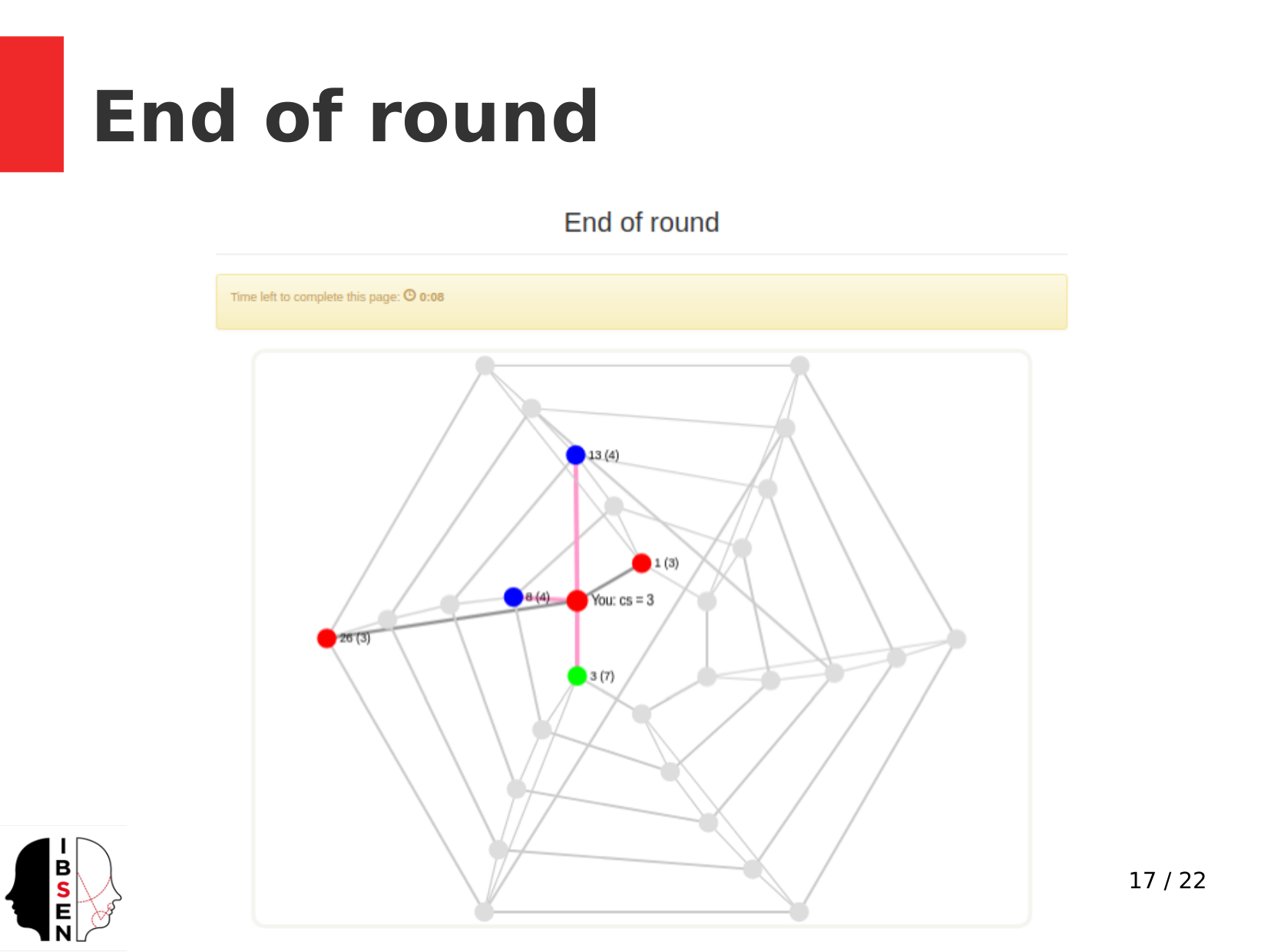}}
\fbox{\includegraphics[page=18,width=0.45\textwidth] {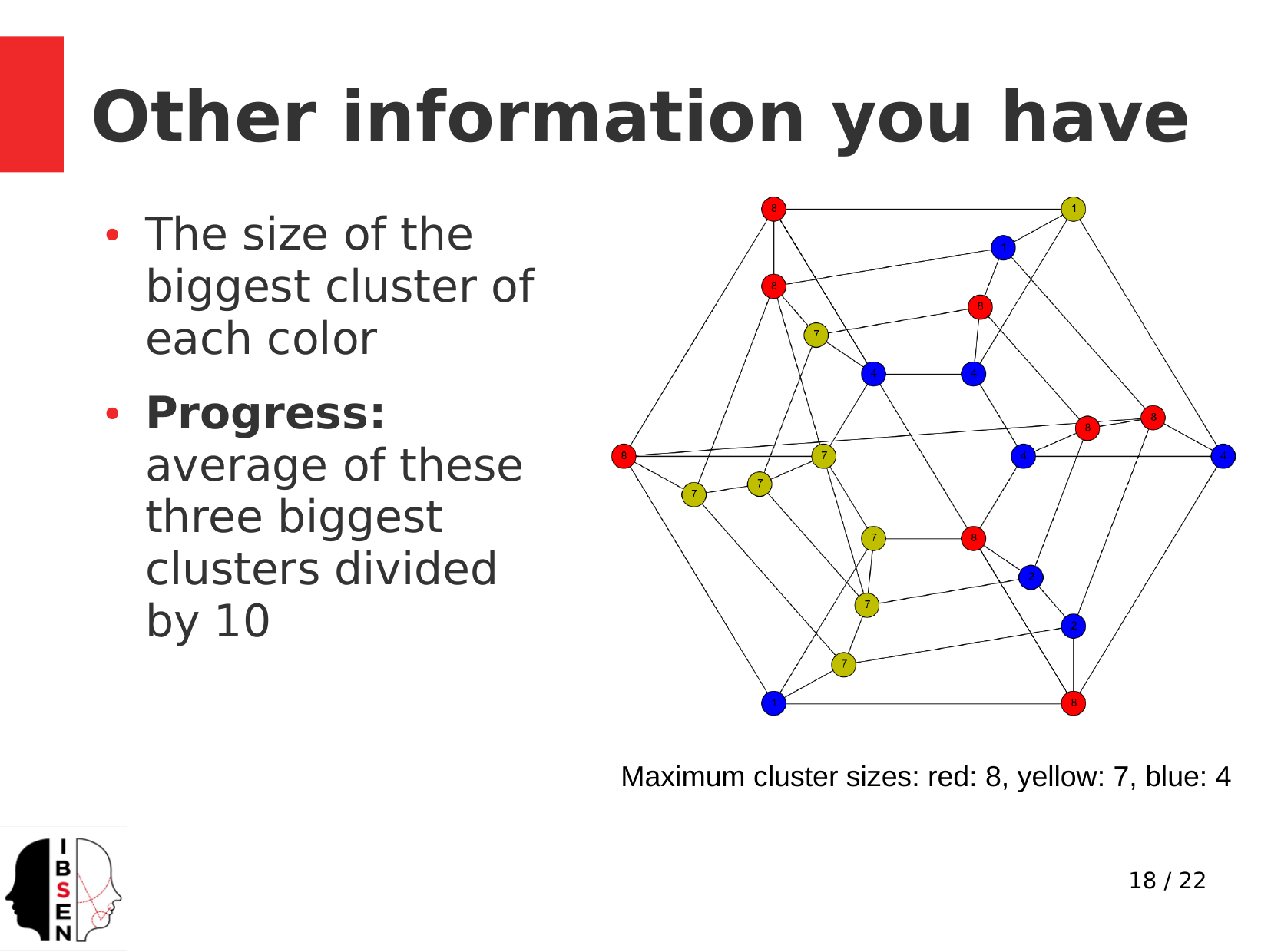}}
\caption{Slides 19-22 of the visual material (presentation) given to the participants. }
\label{fig:Pres3}
\end{figure*}
\clearpage
\newpage

\begin{figure*}[h!]
\centering
\fbox{\includegraphics[page=19,width=0.45\textwidth] {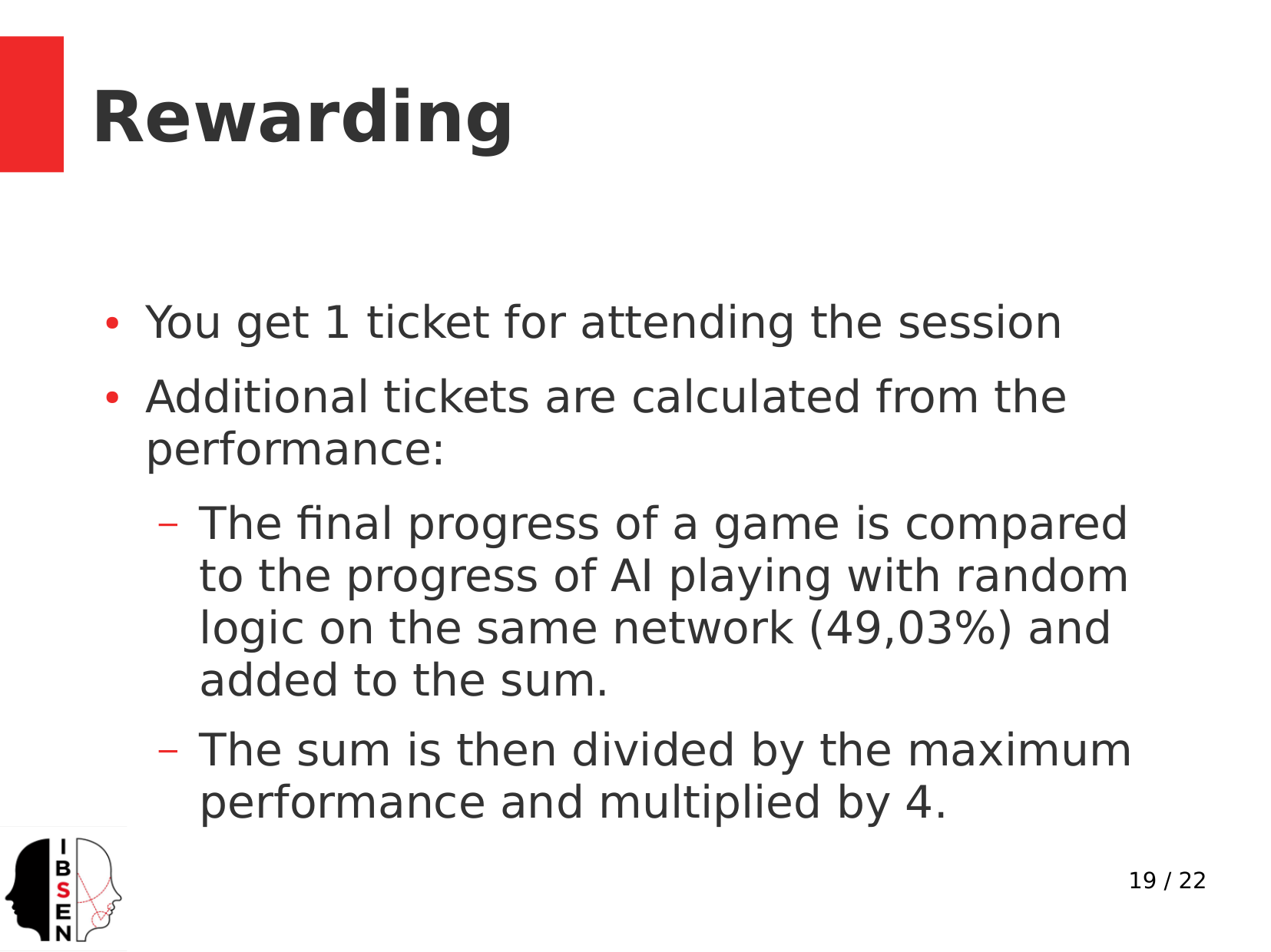}}
\fbox{\includegraphics[page=20,width=0.45\textwidth] {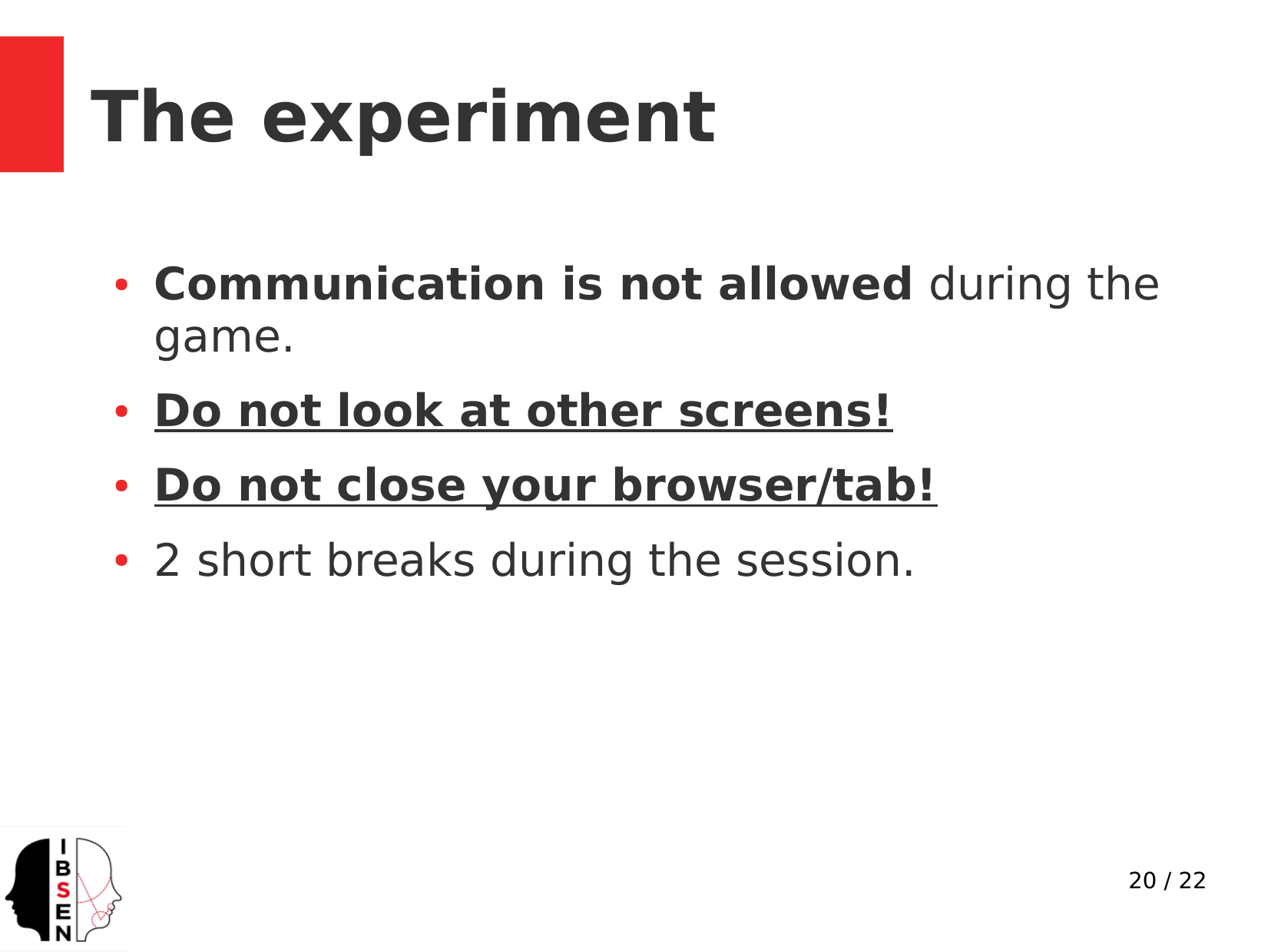}}
\fbox{\includegraphics[page=21,width=0.45\textwidth] {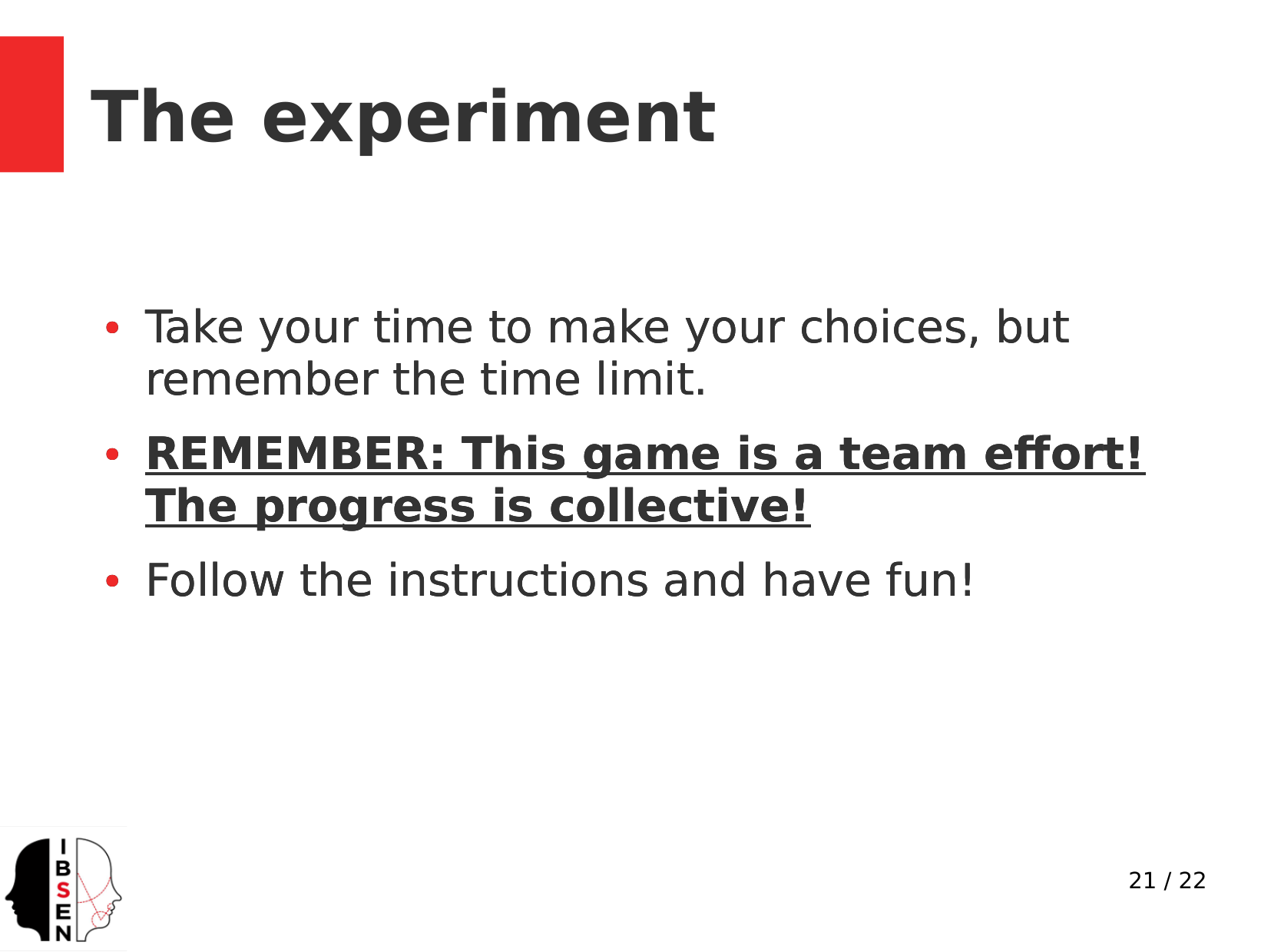}}
\fbox{\includegraphics[page=22,width=0.45\textwidth] {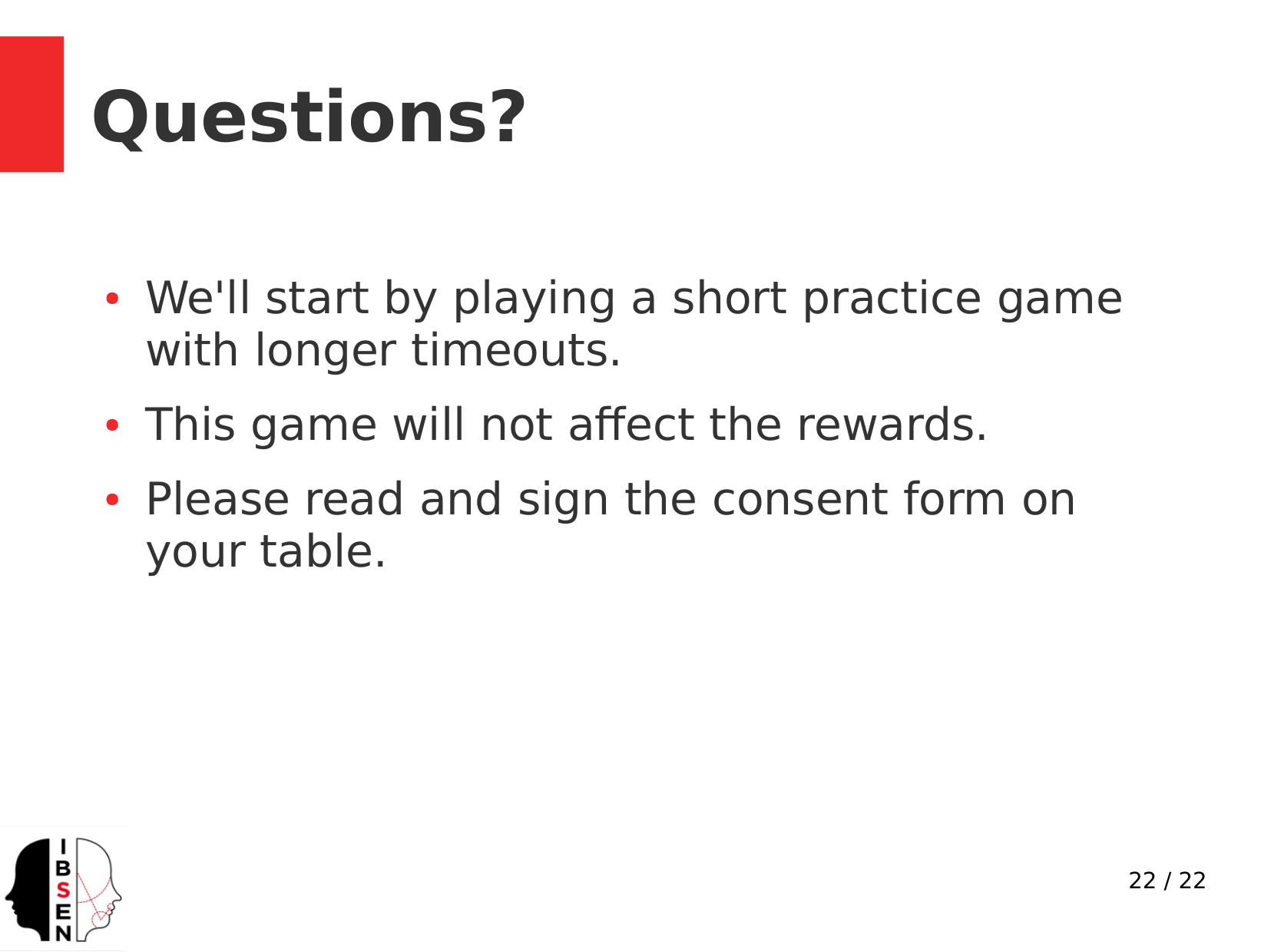}}
\caption{Slides 7-12 of the visual material (presentation) given to the participants. }
\label{fig:Pres4}
\end{figure*}

\end{document}